\documentclass[a4paper,11pt]{article}
\pdfoutput=1 

\usepackage{jheppub} 

\usepackage[T1]{fontenc} 
\usepackage[latin9]{inputenc}
\usepackage{amssymb}
\usepackage{graphicx,ulem,color}
\usepackage{overpic}
\graphicspath{{Images/}}
\usepackage{appendix}
\usepackage{braket}

\title{\boldmath Non-equilibrium dynamics of Goldstone excitation from holography}

\author[a]{Pei Zheng,}
\author[b]{Yidian Chen,}
\author[c]{Danning Li,}
\author[d]{Mei Huang}
\author[a,e]{and Yuxin Liu}

\affiliation[a]{Department of Physics and State Key Laboratory of Nuclear Physics and Technology, Peking University, Beijing 100871, P. R. China}
\affiliation[b]{School of Physics, Hangzhou Normal University, Hangzhou, 311121, P.R. China}
\affiliation[c]{Department of Physics and Siyuan Laboratory, Jinan University,\\601 Huangpu Avenue West, Guangzhou
	510632, P.R. China}
\affiliation[d]{School of Nuclear Science and Technology, University of Chinese Academy
	of Sciences,\\No.19(A) Yuquan Road, Shijingshan District, Beijing 100049, P.R. China}
\affiliation[e]{Center for High Energy Physics, Peking University, Beijing 100871, P. R. China}

\emailAdd{zhengp@stu.pku.edu.cn}
\emailAdd{chenyidian@hznu.edu.cn}
\emailAdd{lidanning@jnu.edu.cn}
\emailAdd{huangmei@ucas.ac.cn}
\emailAdd{yxliu@pku.edu.cn}

\abstract{
By using the holographic approach, we investigate the interplay between the order parameter and Goldstone modes in the real-time dynamics of the chiral phase transition.  By quenching the system to a different thermal bath and obtaining different kinds of initial states, we solve the real-time evolution of the system numerically. Our main focus is on studying far-from equilibrium dynamics of strongly-coupled system and universal scaling behaviors related to such dynamics. The most striking observation is that an additional prethermalization stage emerges at non-critical temperature after introducing the Goldstone modes, which is not reported in any previous studies. Some basic properties related to this additional prethermalization stage have been discussed in detail. More interestingly, we also report a new scaling relation describing non-equilibrium evolution at non-critical temperature. This additional universal behavior indicates the appearance of a non-thermal fixed point in the dynamical region.}

\keywords{Gauge/gravity duality, Prethermalization, Universality, Non-thermal fixed point}

\begin{document} 
\maketitle
\flushbottom

\section{Introduction}
Time is probably the most intriguing parameter in nature. The interests of studying the real-time evolution range from the high-energy physics to condensed matter physics. Prominent examples include the evolution of the early universe shortly after the inflation stage \cite{Kofman:1994rk,Kofman:1997yn,Felder:2000hj,Micha:2002ey}, which refers to a period of strongly accelerated expansion, the unusual initial states created in the collisions of ultra-relativistic nuclei \cite{Baier:2000sb,Berges:2014yta,BRAHMS:2004adc} as well as degenerate quantum gases far from equilibrium \cite{RevModPhys.80.885}. Remarkably, very similar dynamical properties arise in these systems, in spite of the vastly different typical energy scales. In order to understand such universality, a schematic study of real-time evolution of these systems is required.

Typical quantum systems can be classified into two major types: isolated and open systems. The influence of the environment is largely unavoidable in open quantum systems. On the contrary, the dynamics of isolated systems is governed by unitary time evolution and the corresponding effects of environment can be simply neglected. Apart from some particular condensed matter systems, the major investigations of non-equilibrium dynamics are based on isolated quantum system. The typical real-time evolution of isolated systems have been illustrated by J. Berges \textit{et al}. in \cite{Aarts:2001qa,Aarts:2001yn,Berges:2015kfa}. Starting from non-equilibrium, the system can be attracted towards some non-thermal fixed point before ultimately relaxing to thermal equilibrium. 

It is a rather non-trivial question whether a general system can evolve into thermal equilibrium from some non-equilibrium initial states. A quantum system in thermal equilibrium is independent of its initial information and can be characterized by some macroscopic quantities such as conserved charges of the system. It suggests that any thermalization process should involve an effective loss of details of initial conditions. The data from ultra-relativistic heavy-ion collision experiments seem to indicate early thermalization. It is important to realize that different quantities can effectively thermalize on different time scale \cite{Berges:2008wm,Berges:2016nru}. This implies some intermediate stage, with an effective partial memory loss of the initial states, could arise. This corresponding prethermalization can be characterized by some quantities which evolve quasi-stationary, even though the whole system is still out-of-equilibrium. By studying such intermediate stage, possible universality could be discovered.          

When dealing with out-of-equilibrium calculations, there exist some additional complications \cite{kamenev2023field,Calzetta:2008iqa,Berges:2015kfa}. Traditional perturbative scheme relies on expansion of small parameters and has a great success in vacuum and thermal equilibrium. But when switching to the real-time evolution, the perturbative expansion suffers from the so-called secular terms, which typically grow with time and gradually invalidate the expansion approximation even with rather small parameters. In order to resolve such difficulty, infinite summation of perturbative orders is required. Moreover, the universality, which means the insensitivity of the long-time behavior to the detail of the initial information, is expected to encoded in the necessary nonlinear dynamics. One of the commonly used methods suitable for non-equilibrium evolution is based on non-perturbative functional path integral. By using functional path integral, the so-called n-particle irreducible effective action has been widely studied and provides very powerful approximation schemes \cite{Aarts:2002dj,Berges:2004pu,Berges:2001fi}. 

The other powerful tool beyond perturbative expansion is based on gauge/gravity duality \cite{Maldacena:1997re,Witten:1998qj,Gubser:1998bc}. The applications of such frame in the study of bulk properties of strongly coupled plasma has achieved great success. Remarkably, it predicts a celebrated lower bound of shear viscosity  over the volume density of entropy $\eta/s\leq 1/4\pi$ \cite{Kovtun:2004de,Buchel:2003tz,Iqbal:2008by}. Although major achievements of holographic duality are based on linear response theory and are typically valid for those systems near equilibrium, this tool can also be generalized to study far-from-equilibrium dynamic processes. Combining numerical methods in classical numerical relativity, the holographic duality has been successfully applied to study the thermalization of different many-body systems \cite{Balasubramanian:2011ur,Ishii:2015gia,Galante:2012pv,Balasubramanian:2010ce}, the dynamics of phase transition \cite{Attems:2019yqn,Bea:2021zsu,Chen:2022tfy,Chen:2022cwi,Zhao:2023ffs,Bigazzi:2020phm,Bigazzi:2020avc,Bigazzi:2021ucw}, and the extended hydrodynamics \cite{Baggioli:2023tlc,Glorioso:2018mmw,Bhattacharyya:2007vjd,deBoer:2015ija}.

In our work, we use QCD matter system as an example to study related non-equilibrium physics within the holographic framework. The main purposes of this paper are trying to understand the 
typical prethermalization behavior and long-time thermalization behavior of the system involving spontaneous symmetry breaking. In this particular case, the interplay between order parameter and the corresponding Goldstone modes could possibly introduce new phenomena into non-equilibrium dynamics. For this reason, the improved version \cite{Fang:2016nfj} of the soft-wall AdS/QCD model proposed in \cite{Karch:2006pv}, which provides a good framework with consistent description of spontaneous symmetry breaking and Goldstone modes, is our basic starting point. The chiral phase transition \cite{Chelabi:2015gpc,Chelabi:2015cwn} and spectrum of Goldstone modes \cite{Cao:2021tcr,Cao:2022csq} encoded in this model have been previously studied.

This paper is organised as follows. In section \ref{sec:model}, we introduce necessary information about the soft-wall AdS/QCD model. In particular, some subtleties related to different realizations and the corresponding dynamical evolution scheme are discussed. Next in section \ref{sec:non-equilibrium dynamics} we present the major numerical results and the related discussions about prethermalization and universal scaling behaviors related to fixed points. Finally in section \ref{sec:conclusion}, conclusions and further discussions are given.     

\section{Model setup}
\label{sec:model}

\subsection{A brief review of the soft-wall AdS/QCD model}
\label{subsec:soft-wall}
In the holographic framework, from the bottom-up perspective, the hard-wall \cite{Erlich:2005qh} and soft-wall \cite{Karch:2006pv} AdS/QCD models provide an ideal starting point to study some important problems in QCD. By promoting the 4D global $SU(N_f)_L\times SU(N_f)_R$ symmetry of QCD to a gauge symmetry in 5D, these models are very suitable to study those  problems which are related to the spontaneous symmetry breaking of chiral symmetry. To be more explicit, the chiral symmetry is spontaneously broken to $SU(N_f)_V$ when the bulk scalar field gets a non-vanishing expectation value. As a result, the pions appears as the exact Goldstone modes in the chiral limit. Thus, the interplay between Goldstone modes and phase transition can be studied self-consistently by using such models.

In the hard-wall model, the prediction of the spectrum of modes with high excitation number, $n\gg1$, is $m^2_n\propto n^2$, which is not correct compared with QCD data. The soft-wall model extends the hard-wall model by replacing the hard cutoff with a dilaton field $\Phi(z)$, which depends on the fifth dimension $z$. By using soft-wall model, correct mass spectrum of highly excited modes, i.e. the Regge behavior, can be reproduced. So in this paper, we only focus on the soft-wall model.   

The full action in this model is
\begin{eqnarray}
\label{eqn2.1}
    S_{tot}&=& S_G+S_M=\frac{1}{2\kappa^2}\int d^5x\sqrt{-g}e^{-\Phi(z)}(R-2\Lambda)\nonumber\\
    &-&\int d^5x \sqrt{-g}e^{-\Phi(z)}{\rm Tr}\left[|D_M X|^2+V(|X|)+\dfrac{1}{4g_5^2}\left(F_{L,MN}^2+F_{R,MN}^2\right)\right],
\end{eqnarray}
with $g$ the determinant of 5D metric $g_{MN}$, $g_5$ the gauge coupling. $X$ is a bulk matrix-valued scalar field dual to the 4D operator $\bar{\psi}^{\alpha}\psi^{\beta}$ (with $\alpha$ and $\beta$ flavor indices). The corresponding covariant derivative $D_MX$ and gauge field strength $F_{L/R}^{MN}$ are defined as
\begin{eqnarray}
    D_MX&=&\partial_MX-iA_{L,M}X+iXA_{R,M}\label{eqn2.2},\\
    F_{L/R}^{MN}&=&\partial^MA_{L/R}^N-\partial^NA_{L/R}^M-i[A_{L/R}^M,A_{L/R}^N]\label{eqn2.3}.
\end{eqnarray}
For the convenience of later analysis, the gauge fields can also be decomposed into vector and axial-vector fields,
\begin{eqnarray}
    A_{L,M}=V_M+A_M,\qquad A_{R,M}=V_M-A_M\label{eqn2.4},
\end{eqnarray}
then the covariant derivative and gauge field strength can be rewritten as
\begin{eqnarray}
    D_MX&=&\partial_MX-i[V_M,X]-i\{A_M,X\}\label{eqn2.5},\\
    F_V^{MN}&=&\partial^MV^N-\partial^NV^M-i[V^M,V^N]-i[A^M,A^N]\label{eqn2.6},\\
    F_A^{MN}&=&\partial^MA^N-\partial^NA^M-i[V^M,A^N]-i[A^M,V^N]\label{eqn2.7}.
\end{eqnarray}
Finally, by mapping the holographic calculation for the two-point function of the vector current to the 4D field theory calculation, one could fix the gauge coupling $g^2_5=12\pi^2/N_c$ \cite{Erlich:2005qh,Son:2003et}. In this work, the realistic case with $N_c=3$ will be emphasized. 

We do not want to study fully coupled system in this work. Instead, we take flavor part as a probe and neglect back-reaction of flavor part to gravity part. Under such approximation, the dilaton field $\Phi(z)$ and 5D metric $g_{MN}$ should be treated as background fields which must be fixed without involving any dynamics. 

To study thermal behavior related to phase transition, finite temperature effects should be taken into account. So we have the AdS-Schwarzchild black hole solution as our background geometry,
\begin{eqnarray}
   ds^2=e^{2A(z)} \left[-\dfrac{1}{f(z)}dz^2+f(z)dt^2-dx^idx_i\right]\label{eqn2.8},
\end{eqnarray}
with
\begin{eqnarray}
   A(z)=-\ln{z} \ \ {\rm{and}} \ \ f(z)=1-\dfrac{z^4}{z^4_h}\label{eqn2.9},
\end{eqnarray}
where $z_h$ is the horizon defined by $f(z_h)=0$. The temperature of 4D system is just identified as the corresponding Hawking temperature
\begin{eqnarray}
   T=\Big{|} \dfrac{f^{\prime}(z_h)}{4\pi} \Big{|}=\dfrac{1}{\pi z_h}\label{eqn2.10}.
\end{eqnarray}

In this work, we choose the dilaton field profile as quadratic form,
\begin{eqnarray}
   \Phi(z)=\mu^2_g z^2\label{eqn2.11},
\end{eqnarray}
which introduces some IR scale and breaks the conformal symmetry of the dual field theory. Here, $\mu_g$ is a parameter related to some low energy scale which is responsible for the hadron mass spectrum. 

One can only keep mass term in the scalar potential which reads $V(|X|)=m^2_5|X|^2$. According to the AdS/CFT dictionary, the five dimensional mass of scalar field $X$ is $M^2_5=(\Delta-p)(\Delta+p-4)=-3$ by taking $p=0$ and $\Delta=3$. But as pointed out in Ref. \cite{Chelabi:2015gpc,Chelabi:2015cwn}, the above model with $\Phi(z)=\mu^2_g z^2$ and $m^2_5=-3$, can not describe correct spontaneous chiral symmetry breaking. On the contrary, in chiral limit, it gives a vanishing chiral condensate in whole temperature region. According to the study in Ref. \cite{Chen:2018msc}, the scalar mass should be modified to guarantee that the mass squared of the scalar meson vanishes at some critical temperature $T_c$. Besides, higher order corrections in the scalar potential are required to give correct meson spectra and low temperature chiral condensate. Combining all the ingredients together, the final form of scalar potential in this work reads
\begin{eqnarray}
   V(|X|)=m^2_5(z)|X|^2+\lambda |X|^4\label{eqn2.12},
\end{eqnarray}
where the 5D mass square $m^2_5(z)$ takes the following form 
\begin{eqnarray}
m^2_5(z)=-3-\mu^2_c z^2\label{eqn2.13}, 
\end{eqnarray}
as in Ref.\cite{Fang:2016nfj}.

So far, there remains three parameters, $\mu_c$, $\mu_g$, $\lambda$, in our model. These parameters can be fitted by comparing with experimental data for light mesons, the final values of these parameters could be set as \cite{Fang:2016nfj}
\begin{eqnarray}
\label{eqn2.14}
   \mu_g=440{\rm{MeV}},\ \ \mu_c=1450{\rm{MeV}},\ \ \lambda=80.
\end{eqnarray}
With this set of parameters, one can show the presence of exact massless modes in the chiral limit \cite{Cao:2022csq,Cao:2021tcr}, confirming the desired Goldstone nature within this model.

To simplify possible phase structure of the theory, we focus on $N_f=2$ case and take degenerate quark mass which can be denoted as $m_q=m_u=m_d$. When only physics of phase transition is under consideration, we can keep order parameter and neglect all other possible channels. To be more precise, we parameterize $X$ field as
\begin{eqnarray}
   X=\dfrac{\chi}{2}\mathbf{I}_{2\times2}\label{eqn2.15}.
\end{eqnarray}
The field $\chi$ is related to the chiral condensate. Then we can insert eqs. (\ref{eqn2.8}) and (\ref{eqn2.11})-(\ref{eqn2.15}) directly into eq. (\ref{eqn2.1}), the equation of motion can be derived as
\begin{eqnarray}
   \label{eqn2.16}
   \chi^{\prime\prime}+\left(3A^{\prime}+\dfrac{f^{\prime}}{f}-\Phi^{\prime}\right)\chi^{\prime}+\dfrac{e^{2A}}{f}\left[(3+\mu^2_c z^2)-\dfrac{\lambda}{2}\chi^2\right]\chi=0.
\end{eqnarray}
At the UV boundary, the asymptotic expansion of $\chi$ field simply reads
\begin{eqnarray}
   \label{eqn2.17}
   \chi(z\rightarrow0)=m_q\zeta z+\dfrac{\sigma}{\zeta}z^3+\dfrac{m_q\zeta}{4}(4\mu^2_g-2\mu^2_c+m^2_q\zeta^2\lambda)z^3\ln{z}+\mathcal{O}(z^4),
\end{eqnarray}
where $m_q$ and $\sigma$ are interpreted as quark mass and chiral condensate, respectively, by using the holographic dictionary. The normalization factor $\zeta=\sqrt{N_c}/2\pi$ is introduced by comparing the two-point function of $\braket{\sigma(k)\sigma(0)}$ with the result from 4D calculation in large momentum limit \cite{Cherman:2008eh}.

\begin{figure}[ht]
    \centering
    \includegraphics[scale=0.3]{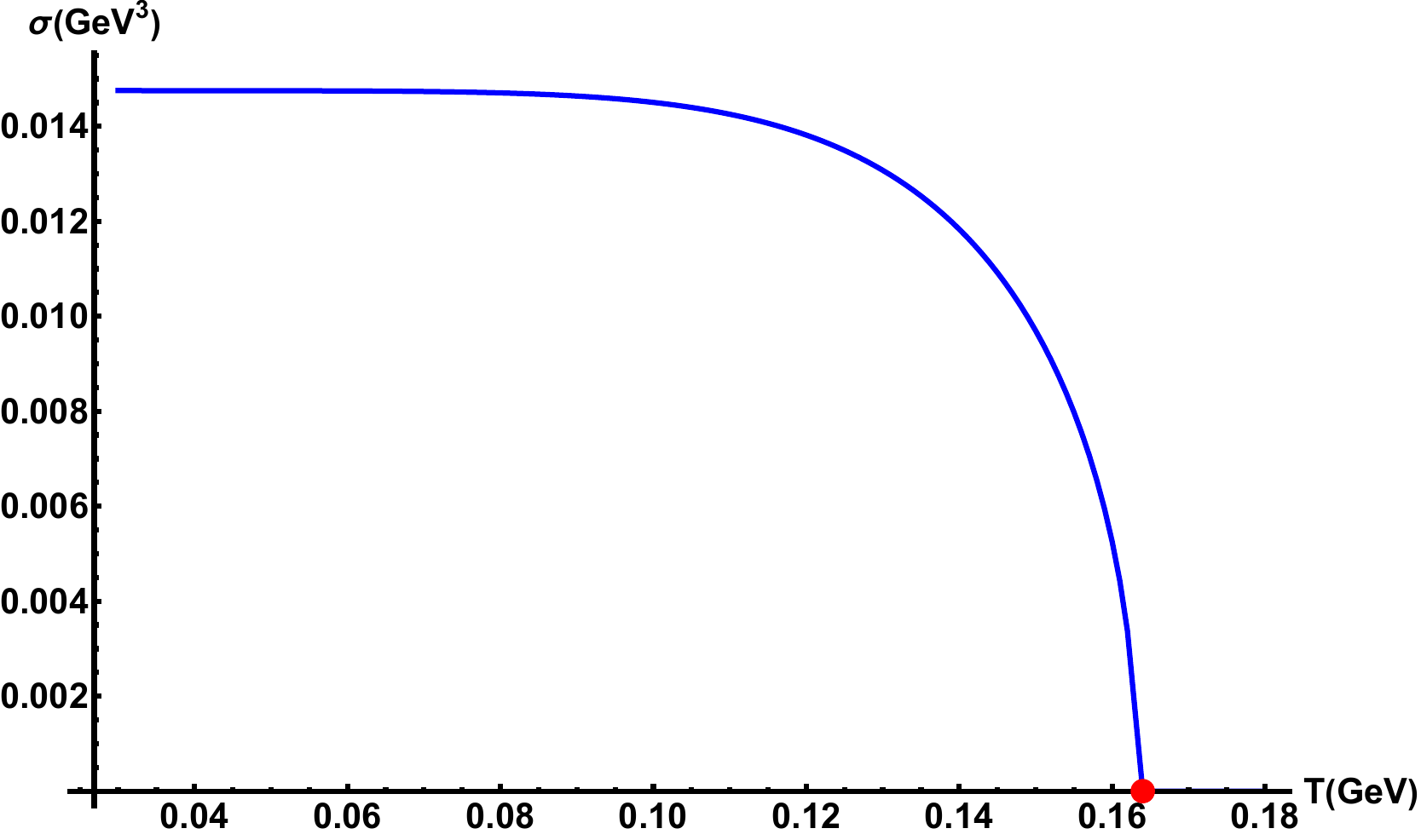}
   \caption{Chiral condensation $\sigma$ as a function of temperature $T$ in the chiral limit. The red circle represents the critical point ($T_c$) of the second order phase transition. Numerically, by using the parameters (\ref{eqn2.14}), we get $T_c=163{\rm{MeV}}$.}
   \label{fig:phase}
\end{figure}

By solving eq. (\ref{eqn2.16}) numerically, we can extract the information about chiral condensate which can be used to characterize phase transition behavior. Due to the need of Goldstone modes in later discussion, here we only focus on the chiral limit. In Fig. \ref{fig:phase}, we show the chiral phase transition in the desired chiral limit. In this case, the chiral phase transition is a second-order phase transition with critical temperature $T_c=163{\rm{MeV}}$. The following discussion involving Goldstone modes will heavily rely on this particular phase diagram. 

\subsection{Linear realization vs. nonlinear realization}
\label{subsec:comparison of realization}
In this work, we want to study interplay between order parameters and possible Goldstone modes. In addition to the previous studied $\chi$ field, which refers to order parameter only, we should introduce more degrees of freedom in our model. Note that the $X$ field in our original model is a matrix-valued scalar field with two flavor indices attached to it. Similar to the method used in ordinary chiral Lagrangian, we use the group structure of $X$ field to parametrize it appropriately. The ambiguity of parameterization remains in the present setup. In this section, we briefly discuss this problem by comparing two typical parametrization schemes.     

The most straightforward way is using so-called linear realization which parametrizes $X$ field as
\begin{eqnarray}
   \label{eqn2.18}
   X=\dfrac{1}{2}(\chi \mathbf{I}_{2\times2}+i \pi^a\tau^a),
\end{eqnarray}
where $\tau^a$ and $\pi^a$ refer to the usual Pauli matrices and pseudoscalar fields, respectively. The normalization factor $1/2$ here is introduced to make kinetic terms canonical in the action. In this case, by using holographic dictionary, these pseudoscalar fields are dual to desired Goldstone modes. We do not want to discuss some specific problems related to hadron physics, so we will treat these fields as Goldstone modes in the rest part of this paper instead of identifying these as pions.

Turning our attention to the gauge fields sector, vector fields are irrelevant to our present problem, so we simply ignore vector channel in our action. 

Inserting ansatz (\ref{eqn2.18}) into flavor part of the action (\ref{eqn2.1}), and neglecting all the terms involving vector fields, we can get the corresponding effective action as
\begin{align}
   \label{eqn2.19}
   S_{{\rm{eff}}}=&-\int d^5x \sqrt{-g}e^{-\Phi}\left\{\dfrac{1}{2}\partial_M\chi\partial^M\chi+\dfrac{1}{2}\partial_M\vec{\pi}\cdot\partial^M \vec{\pi}+(\partial_M\chi)\vec{A}^M\cdot \vec{\pi}-\dfrac{1}{2}\chi(\partial_M\vec{\pi})\cdot\vec{A}^M\right.\nonumber\\
&\left.+\dfrac{1}{8}\chi^2\vec{A}_M\cdot\vec{A}^M+\dfrac{1}{2}(\vec{A}_M\cdot\vec{\pi})(\vec{A}^M\cdot\vec{\pi})+\dfrac{1}{2}m^2_5\left(\chi^2+\vec{\pi}\cdot\vec{\pi}\right)+\dfrac{1}{8}\lambda\left(\chi^2+\vec{\pi}\cdot\vec{\pi}\right)^2\right.\nonumber\\
&\left.+\dfrac{1}{4g^2_5}\left[(\vec{A}_M\cdot\vec{A}^M)(\vec{A}_N\cdot\vec{A}^N)-(\vec{A}_M\cdot\vec{A}_N)(\vec{A}^M\cdot\vec{A}^N)+F^{MN,a}F^a_{MN}\right]\right\},
\end{align}
where the arrow refers to isospin vector and dot product represents the summation of isospin indices. When writing down the form of effective action (\ref{eqn2.19}), we define a strength tensor in a common Abelian fashion but with a non-Abelian index attached to it
\begin{eqnarray}
\label{eqn2.20}
    F^{MN,a}\equiv\partial^MA^{Na}-\partial^NA^{Ma}.
\end{eqnarray}
Note the derivation of the effective action (\ref{eqn2.19}) does not rely on any perturbative expansion procedure. Different from usual framework which treats $\pi^a$ fields as perturbation, we put all the fields on an equal footing. Besides, for the future convenience, we maintain the covariance formulation of effective action without specifying any particular background geometry. 

Another commonly used parameterization is the nonlinear one. In this nonlinear realization, the matrix field $X$ can be parametrized as 
\begin{eqnarray}
\label{eqn2.21}
    X=\dfrac{\chi}{2}e^{i\varphi^a\tau^a},
\end{eqnarray}
where $\varphi^a$ fields are the phase in isospin space. In this work, we take the same convention as \cite{Grossi:2020ezz,Grossi:2021gqi} and define 
\begin{eqnarray}
    \label{eqn2.22}
    \varphi^a=\dfrac{\pi^a}{\braket{\sigma}}.
\end{eqnarray}
Specific to the  $SU(2)$ case, the exponent can be further expressed as  
\begin{eqnarray}
\label{eqn2.23}
    X=\dfrac{\chi}{2}\left(\cos{\varphi}\mathbf{I}_{2\times2}+i\sin{\varphi}(n^a\tau^a)\right)
\end{eqnarray}    
where $\varphi=\sqrt{\varphi^a\varphi^a}$ and $n^a=\varphi^a/\varphi$.

With the help of eq. (\ref{eqn2.23}), all the commutators or anti-commutators involved in covariant derivatives and field strength can be calculated directly. After tracing all the group indices and neglecting irrelevant contributions from vector channel, the effective action in the nonlinear realization can be written as
\begin{align}
\label{eqn2.24}
S_{{\rm{eff}}}=&-\int d^5x \sqrt{-g}e^{-\Phi}\left\{\dfrac{1}{2}\partial_M\chi\partial^M\chi+\dfrac{1}{2}\chi^2\partial_M\varphi\partial^M\varphi-\chi^2(\partial_M\varphi)(A^{Ma}n^a)\right.\nonumber\\
&\left.+\dfrac{1}{2}\chi^2\sin^2{\varphi}(A^a_MA^{Mb})(n^an^b)+\dfrac{1}{2}\chi^2\cos^2{\varphi}(A^a_MA^{Ma})+\dfrac{1}{2}m^2_5\chi^2+\dfrac{1}{8}\lambda\chi^4\right.\nonumber\\
&\left.+\dfrac{1}{4g^2_5}\left[(\vec{A}_M\cdot\vec{A}^M)(\vec{A}_N\cdot\vec{A}^N)-(\vec{A}_M\cdot\vec{A}_N)(\vec{A}^M\cdot\vec{A}^N)+F^{MN,a}F^a_{MN}\right]\right\},   
\end{align}
where the same conventions and approximations as linear realization have been used. 

Comparing these two kinds of parametrization, very different forms of effective action appear. But in principle, all these realizations should be related because they originate from a single field $X$. In this paper, $SU(2)$ case is the only one under consideration. Accordingly, the relations between these two realizations can be simplified a lot. By matching components along different independent directions in isospin space, we obtain
\begin{eqnarray}
\chi_{l}&=&\chi_{nl}\cos{\varphi}\label{eqn2.25}, \\
\pi^a&=&n^a\chi_{nl}\sin{\varphi}\label{eqn2.26},
\end{eqnarray}
where the subscripts are introduced to distinguish two different realizations. Obviously, there are no preferred direction in eq. (\ref{eqn2.26}), so we can choose any specific direction as representative and drop isospin index safely. In order to extract physical observables, we use near boundary asymptotic expansion and holographic dictionary, which can be written as
\begin{eqnarray}
\chi_l(z\rightarrow 0)&=&\dfrac{\braket{\sigma}_l}{\zeta}z^3+\cdots \label{eqn2.27}, \\
\pi(z\rightarrow 0)&=&\braket{\pi} z^3+\cdots \label{eqn2.28},\\
\chi_{nl}(z\rightarrow 0)&=&\dfrac{\braket{\sigma}_{nl}}{\zeta}z^3+\cdots \label{eqn2.29}, \\
\varphi(z\rightarrow 0)&=&\braket{\varphi}z^0+\cdots\label{eqn2.30}.
\end{eqnarray}
To clarify, the above expansions are only valid in the chiral limit and all possible external sources are vanished. Higher order terms are not needed because we only care about one-point functions. The expected relations can be obtained directly from leading order terms in expansion of eqs. (\ref{eqn2.25}) and (\ref{eqn2.26}). Keeping the definition (\ref{eqn2.22}) in mind, we expand equations mentioned above directly and the final matching relations are given by
\begin{eqnarray}
\braket{\sigma}_l&=&\braket{\sigma}_{nl}\cos\left(\dfrac{\braket{\pi}_{nl}}{\braket{\sigma}_{nl}}\right)\label{eqn2.31},\\
\braket{\pi}_{l}&=&\dfrac{\braket{\sigma}_{nl}}{\zeta}\sin\left(\dfrac{\braket{\pi}_{nl}}{\braket{\sigma}_{nl}}\right)\label{eqn2.32}.
\end{eqnarray}
Note these relations are derived in coordinate space. When trying to Fourier transformed into momentum space, these relations will lead to nonlinear couplings of different momentum modes and can not be formulated as compact algebraic relations.

Furthermore, special attention should be paid to those interaction terms which involve derivative couplings in the effective action (\ref{eqn2.24}). Compared to the effective action derived by using linear realization, these terms will modify the structure of the corresponding equations of motion and bring in extra numerical difficulties when solving equations of motion. We have made great efforts to solve the problem by using nonlinear formulation, but the corresponding numerical solutions suffer from severe numerical instabilities and no physical meaningful result can be extracted at present. It follows that only linear realization will be used in the following study of non-equilibrium dynamics and nonlinear case will be postponed to future work.  

\subsection{Dynamical evolution scheme and corresponding EOM}
\label{subsec:eom}

In order to study appealing non-equilibrium phenomena, we need to prepare the initial states of the system elaborately. Besides, we also need to study dynamical behaviors under real-time evolution. Therefore, in this section, we illustrate our basic scheme suitable for non-equilibrium dynamics and show the brief derivation of the corresponding equations of motion.     

There exist some external parameters which can be used to character the states of our system. During the non-equilibrium evolution, all these parameters can change with time. To simplify the discussion, we use function $R(t)$ to represent all possible external parameters collectively. In this work, we assume that the system will finally evolve into some specific final state and denote it as $R_f=R(t\rightarrow\infty)$. Further, we assume the final states of system are some states in thermal equilibrium. This, certainly, is not a trivial assumption, but all our numerical results are not in contradiction with this assumption and final thermalization can be observed. Thus, we treat the assumptions mentioned above reasonable.       
\begin{figure}[ht]
    \centering
    \includegraphics[scale=0.4]{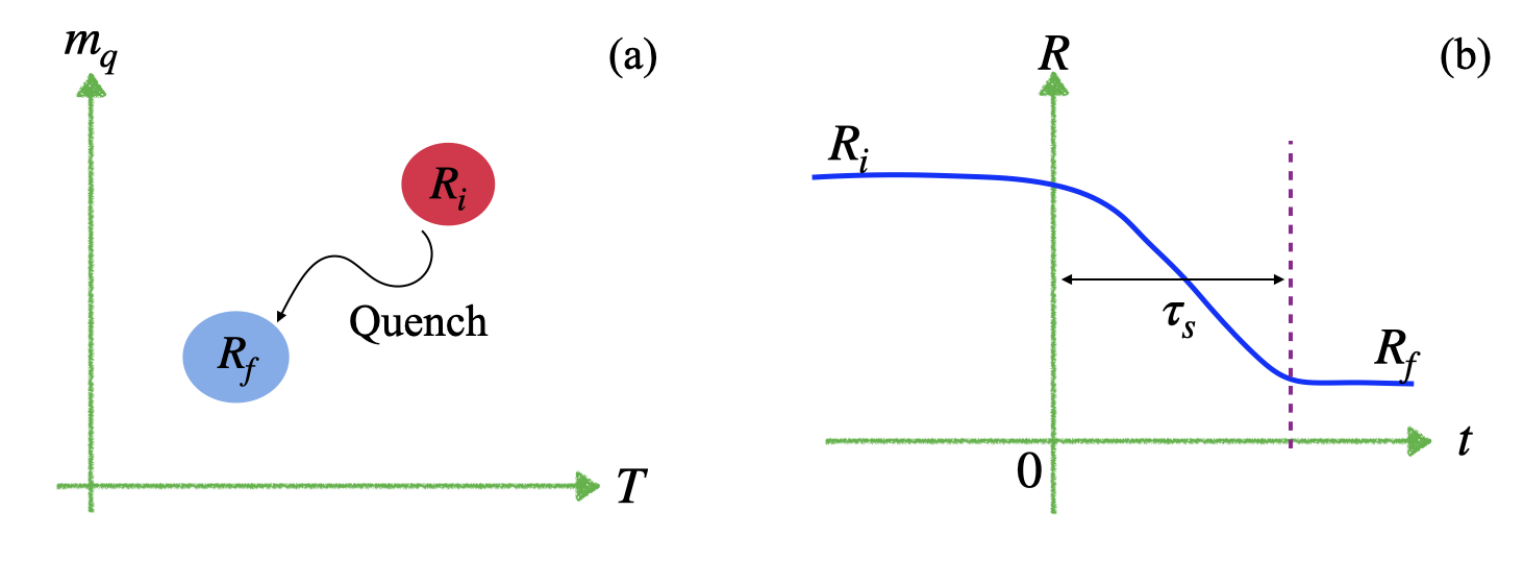}
   \caption{Schematics of the quench procedure. (a) General quench protocols. (b) Time dependence of the external parameters $R(t)$.}
   \label{fig:quench}
\end{figure}

As shown in Fig. \ref{fig:quench}, we start from a given state $R_i(t=t_i)$, and quench from $R_i$ to the state parametrized by $R_f$ within a finite timescale $\tau_s$. To simplify the numerical calculation, we only consider the sudden quench case, i.e., $\tau_s\rightarrow0$, in this work. To clarify, we should write the time dependence of $R(t)$ as 
\begin{align}
    \label{eqn2.33}
    R(t)=R_i+\theta(t) (R_f-R_i),
\end{align}
where $\theta(t)$ is the step function.

After specifying the initial states, we then study the real-time evolution. In general, there exists divergence near the black horizon. To avoid this difficulty, we transform to the Eddington-Finkelstein (EF) coordinates, and now the metric can be written as
\begin{eqnarray}
\label{eqn2.34}
ds^2=\dfrac{1}{z^2}\left[-f(z)dt^2-2dtdz+dx^idx^i\right].
\end{eqnarray}
Here we use $t$ coordinate as EF time without introducing any ambiguity. The other advantage of using EF coordinates is that EF time in bulk theory reduces to physical time in boundary field theory at the boundary $z=0$, but remains non-singular at the horizon. 

Accordingly, with the metric (\ref{eqn2.34}), we derive equations of motion directly from effective action (\ref{eqn2.19}). The first one is a constraint from variation of $A^a_t$ field, which reads
\begin{align}
    \label{eqn2.35}
    \partial_z(\sqrt{-g}e^{-\Phi}(g^{zt})^2\partial_z A^a_t)+\sqrt{-g}e^{-\Phi}g^{zt}g^{ij}(\partial_z\partial_jA^a_i)+g^2_5 \sqrt{-g} e^{-\Phi}g^{zt}(\partial_z\chi)\pi^a\nonumber&\\
    -\dfrac{1}{2}g^2_5\sqrt{-g}e^{-\Phi}g^{zt} \chi (\partial_z \pi^a)&=0.
\end{align}
From holographic dictionary, the $A^a_t$ field is related to the axial charge density and the corresponding chemical potential. In this paper, we do not introduce any conserved charge, and accordingly, there are no non-zero chemical potential and charge density involved in the system. In the chiral limit and with vanishing external sources, one can show that constraint (\ref{eqn2.35}) can be satisfied when inserting near-boundary expansions of all the fields. Therefore, we can safely ignore the contribution from $A^a_t$ field in the system.

The remaining equations of motion which are closely related to dynamics can be written as
\begin{align}
    \partial_N(\sqrt{-g}e^{-\Phi}g^{MN} \partial_M\chi)-\dfrac{1}{4}\sqrt{-g}e^{-\Phi}g^{ij}(A^a_iA^a_j)\chi-\sqrt{-g} e^{-\Phi}m^2_5\chi \nonumber&\\
    -\dfrac{1}{2}\sqrt{-g}e^{-\Phi}\lambda\left(\chi^2+\pi^a\pi^a\right)\chi
    &=0, \label{eqn2.36}\\
    \partial_N(\sqrt{-g}e^{-\Phi}g^{MN} \partial_M\pi^a)-\sqrt{-g}e^{-\Phi}g^{ij}(A^b_i\pi^b)A^a_j-\sqrt{-g} e^{-\Phi}m^2_5\pi^a \nonumber&\\
    -\dfrac{1}{2}\sqrt{-g}e^{-\Phi}\lambda\left(\chi^2+\pi^b\pi^b\right)\pi^a&=0, \label{eqn2.37}\\
    \partial_z(\sqrt{-g}e^{-\Phi}g^{zz}g^{ij}(\partial_zA^a_i))+\partial_z(\sqrt{-g}e^{-\Phi}g^{zt}g^{ij}(\partial_tA^a_i))+\sqrt{-g}e^{-\Phi}g^{zt}g^{ij}(\partial_t\partial_zA^a_i)\nonumber&\\
    -g^2_5 \sqrt{-g}e^{-\Phi}\left[\dfrac{1}{4}\chi^2g^{ij}A^a_i+g^{ij}(A^b_i\pi^b)\pi^a\right]-\sqrt{-g}e^{-\Phi}g^{ij}g^{mn}\left[(A^b_mA^b_n)A^a_i-(A^b_iA^b_m)A^a_n\right]&=0.\label{eqn2.38}
\end{align}
Note eq. (\ref{eqn2.38}) is degenerate with respect to the spatial indices $i$. So we can choose one spatial component of $A^a_i$ as representative and safely ignore the explicit spatial indices $i$. Besides, we use following trick to simplify the group structure
\begin{eqnarray}
  \label{eqn2.39}
  \sum_{i}A^a_iA^b_i=f\delta^{ab}\ \ {\rm{and}} \ \ \pi^a\pi^b=h\delta^{ab},
\end{eqnarray}
because $\delta^{ab}$ is the only symmetric invariant tensor in $SU(2)$ group. Contracting both sides of eq. (\ref{eqn2.39}) with $\delta^{ab}$, one can easily solve the unknown function and get
\begin{eqnarray}
\label{eqn2.40}
  f=\dfrac{1}{3}\sum_{i,a}A^a_iA^a_i \ \ {\rm{and}} \ \ h=\dfrac{1}{3}\sum_a\pi^a\pi^a.
\end{eqnarray}
Due to the unbroken isospin symmetry in our setup, we simply keep only one specific isospin channel and ignore all others. This suggests that the group indices can be dropped in the rest part of this paper. 

According to the Goldstone nature of $\pi$ field, the spatial momentum of such modes can not vanish. In principle, this physical property requires non-trivial spatial coordinate dependence of $pi$ field, but solving equations with extra spatial degree of freedom will lead to some numerical difficulties. Thus, we postpone more rigorous treatment to the future work. In this work, we approximately substitute the origin $\pi$ field with its Fourier transformed counterpart and neglect modes mixing introduced by nonlinear terms in the equations. 

Combine all the approximations discussed before, we get the final form of equations suitable for direct numerical procedure
\begin{align}
    z^2f(z)(\partial^2_z\chi)+\left[z^2f^{\prime}(z)-f(z)(3z+z^2\Phi^{\prime}(z))\right](\partial_z\chi)-\left[2z^2\partial_z-3z-z^2\Phi^{\prime}(z)\right]\partial_t\chi \nonumber&\\
    =\dfrac{1}{4}z^2 A^2\chi+m^2_5\chi+\dfrac{\lambda}{2}\left(\chi^2+\pi^2\right)\chi&,\label{eqn2.41}\\
    z^2f(z)(\partial^2_z\pi)+\left[z^2f^{\prime}(z)-f(z)(3z+z^2\Phi^{\prime}(z))\right](\partial_z\pi)-\left[2z^2\partial_z-3z-z^2\Phi^{\prime}(z)\right]\partial_t\pi \nonumber&\\
    =\dfrac{1}{3}z^2A^2\pi+m^2_5\pi+z^2\boldsymbol{p}^2\pi+\dfrac{\lambda}{2}\left(\chi^2+\pi^2\right)\pi&,\label{eqn2.42} \\
    z^2f(z)(\partial^2_z A)+\left[z^2f^{\prime}(z)-f(z)(z+z^2\Phi^{\prime}(z))\right](\partial_zA)-\left[2z^2\partial_z-z-z^2\Phi^{\prime}(z)\right]\partial_tA \nonumber&\\
    =g^2_5\left(\dfrac{1}{4}\chi^2+\dfrac{1}{3}\pi^2\right)A+\dfrac{2}{3}z^2A^3&,\label{eqn2.43}
\end{align}
where $\pi$ field above should be interpreted as Goldstone mode with specific spatial momentum $\boldsymbol{p}$. Eqs. (\ref{eqn2.41})-(\ref{eqn2.43}) are the desired equations which are solved numerically.  

\section{Non-equilibrium dynamics and numerical results}
\label{sec:non-equilibrium dynamics}
According to the previous study on the non-equilibrium process \cite{Cao:2022mep,Flory:2022uzp}, the typical behavior of dynamical evolution can be divided into three stages. The first stage is associated with the underlying microscopic timescale and is dependent on the initial conditions. The intermediate stage, often referred to as prethermalization, is a universal far-from-equilibrium phenomenon. As evolution progresses, all information about the initial conditions is expected to dissipate, and the system's state gradually approaches thermal equilibrium. Thus, the final stage is characterized by long-term relaxation or thermalization. In this section, we discuss the non-equilibrium dynamics of the system involving the Goldstone channel.

To realize the non-equilibrium evolution, considering the chiral limit and vanishing extra external sources, we can take the asymptotic solutions as the approximation of initial state of system,
\begin{align}
\chi(z,t=t_i)&=\dfrac{\sigma_i}{\zeta}z^3\equiv \sigma_{\rm{ini}} z^3,\label{eqn3.1}  \\
\pi(z,t=t_i)&=\braket{\pi}_iz^3\equiv \pi_{\rm{ini}} z^3.\label{eqn3.2}
\end{align}
Besides, all the numerical results presented below rely on the sudden quench to some specific thermal state with well-defined temperature.

\begin{figure}[ht]
    \centering
    \includegraphics[scale=0.3]{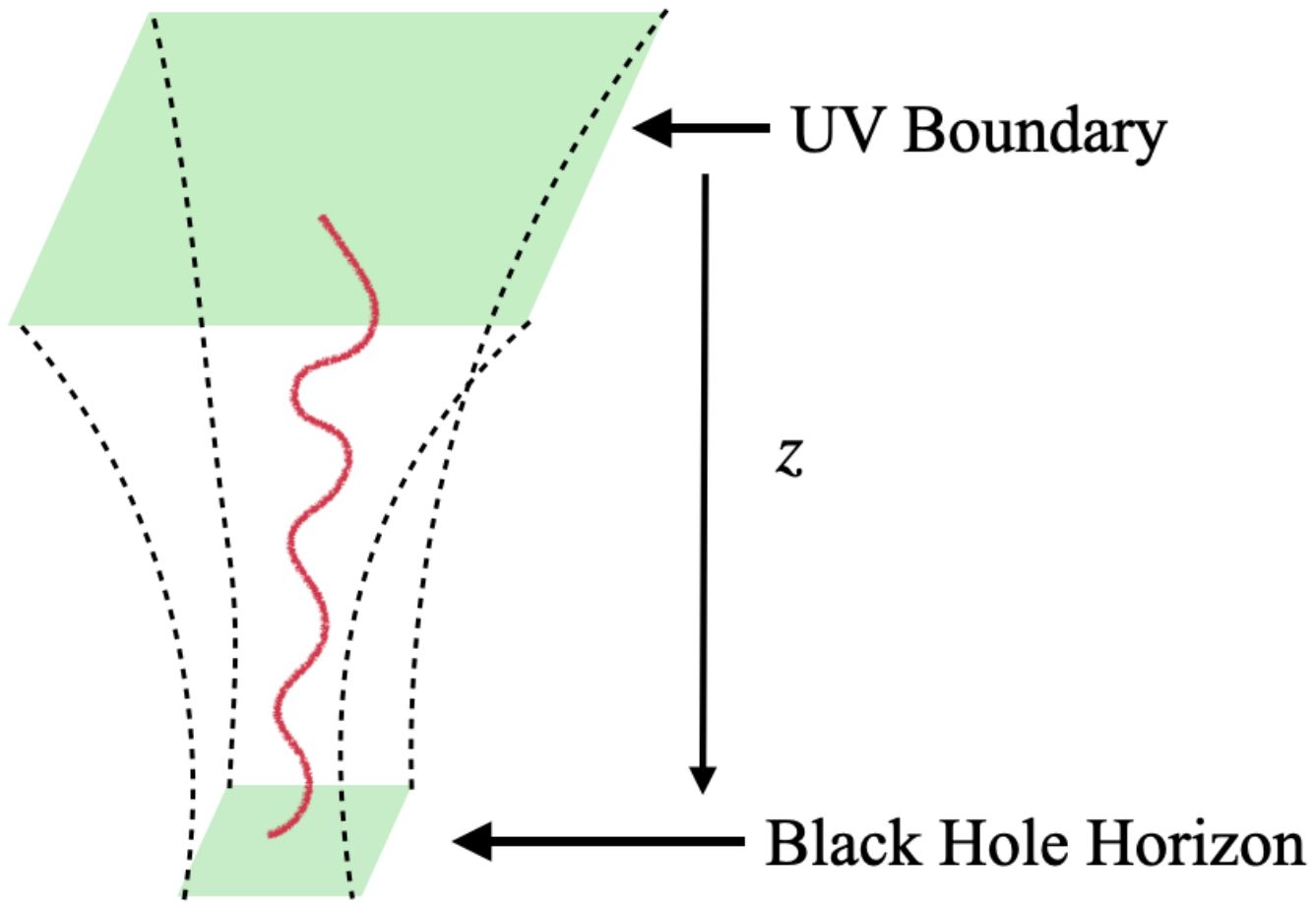}
   \caption{Basic picture of our setup of sudden quench. The red curve simply refers to any bulk field configuration at initial time. With fixed black hole background, these bulk fields can be chosen arbitrarily and refer to some non-equilibrium initial state.}
   \label{fig:basic}
\end{figure}

The basic physical setting described before can be shown in Fig. \ref{fig:basic}. We treat black hole as a big thermal background with fixed temperature and do not disturb it. The bulk fields between UV boundary and black horizon can be disturbed arbitrarily at initial time. This can be interpreted as choosing some non-equilibrium states in boundary field theory by using holographic duality.

\subsection{Prethermalization}
\label{subsec:prethermal}
The origin definition of prethermalization describes the very rapid establishment of an almost constant ratio of pressure over energy density \cite{Berges:2004ce}. It follows that prethermalized quantities can take their equilibrium thermal values with the occupation numbers of corresponding modes deviating from the late-time Bose-Einstein or Fermi-Dirac distribution. This phenomenon occurs on time scales shorter than the time scales related to the final thermalization. In our framework, we do not know how to define occupation number density consistently. Instead, we use order parameter as the representative and study its dynamical evolution behavior arising in the intermediate time. 

\subsubsection{Numerical result at critical temperature}
\label{subsubsec:critical temperature}
\begin{figure}[ht]
    \centering
    \includegraphics[scale=0.3]{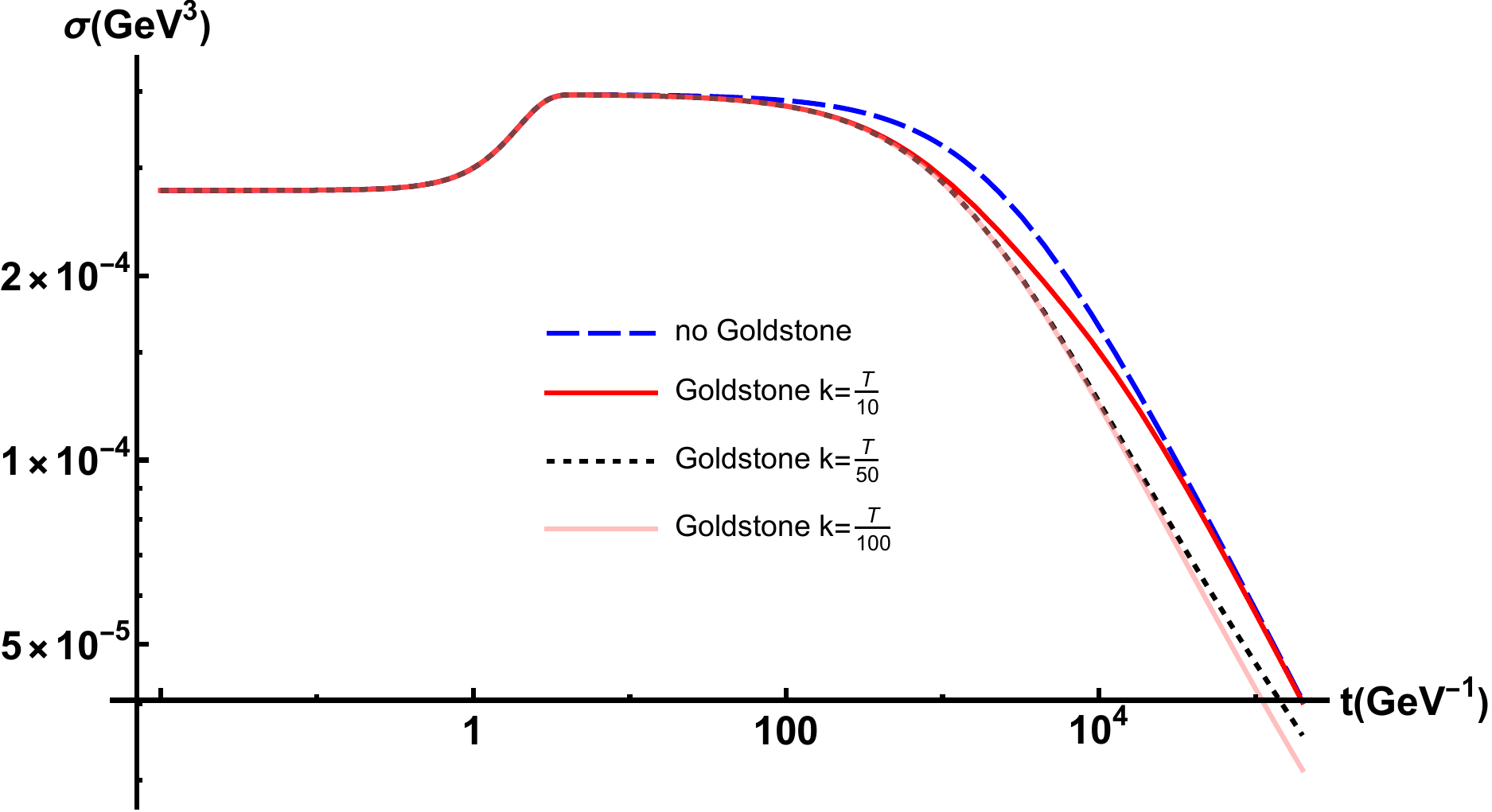}
   \caption{The typical evolution curves of chiral condensate with final thermal state at critical temperature. We compare systems coupling to Goldstone modes with three different spatial momentum to the system with only order parameter involved. The behaviors of plateaus at intermediate timescale are very similar to each other. At later times, clear differences between those cases are found.}
   \label{fig:criticalMom}
\end{figure}

Similar to the previous study, we expect that the prethermalization stage should naturally emerge  when the system suddenly quench to the final thermal state at the critical temperature. In Fig. \ref{fig:criticalMom}, we directly show the comparison between a typical evolution curve without any Goldstone modes and those evolution curves with specific Goldstone mode couplings. Remarkably, the dynamical evolution behaviors involving Goldstone modes exhibit clear evidence of universality associated with prethermalization at the critical point. By introducing Goldstone modes in the system, we do not modify typical dynamical behavior in the full real-time evolution. The general statements of three dynamical stages still hold. However, the concrete quantitative results rely on the Goldstone modes.       

\begin{figure}[htbp]
    \centering
    \begin{overpic}[width=0.45\textwidth]{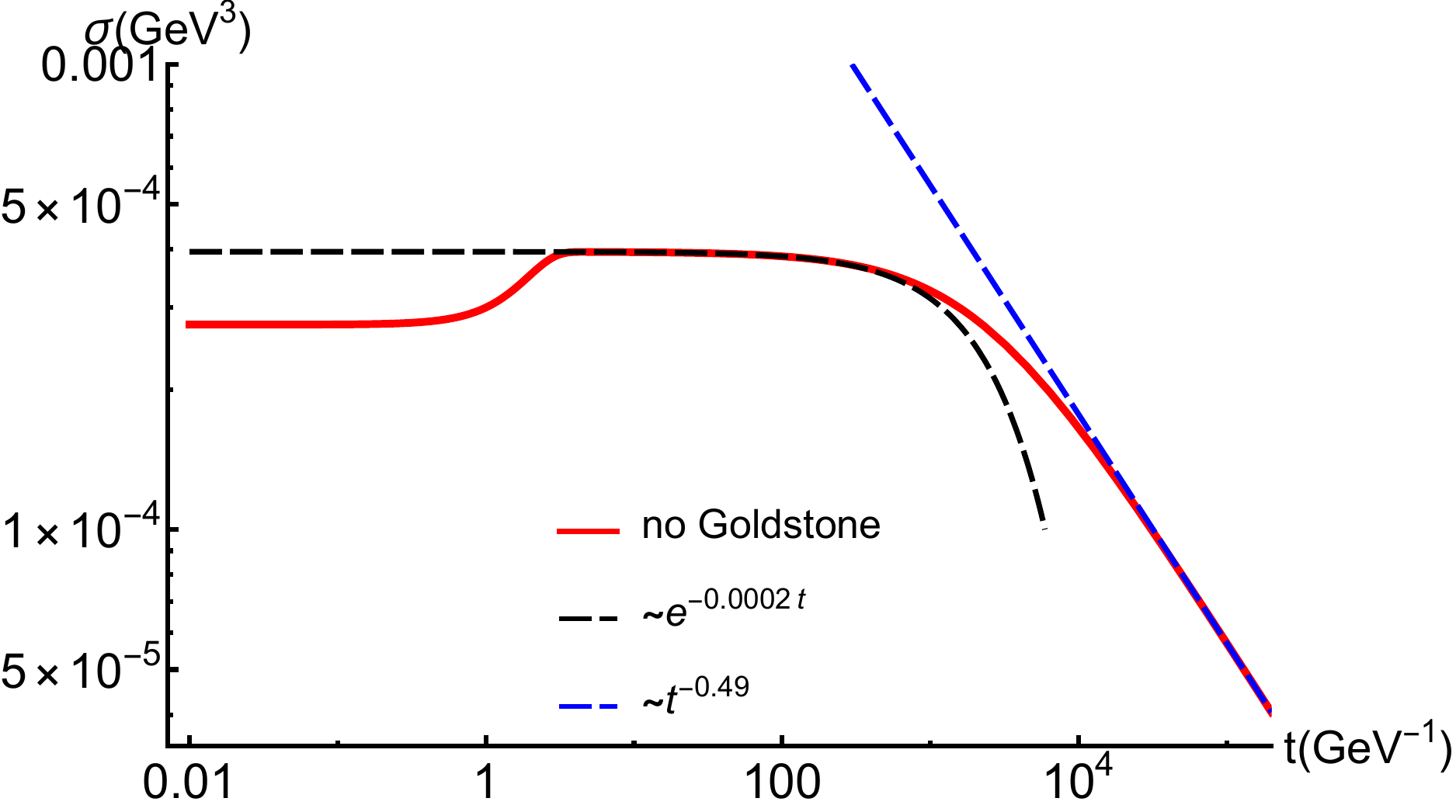}
    \put(20,10){\bf{(a)}}
    \end{overpic}
        \begin{overpic}[width=0.45\textwidth]{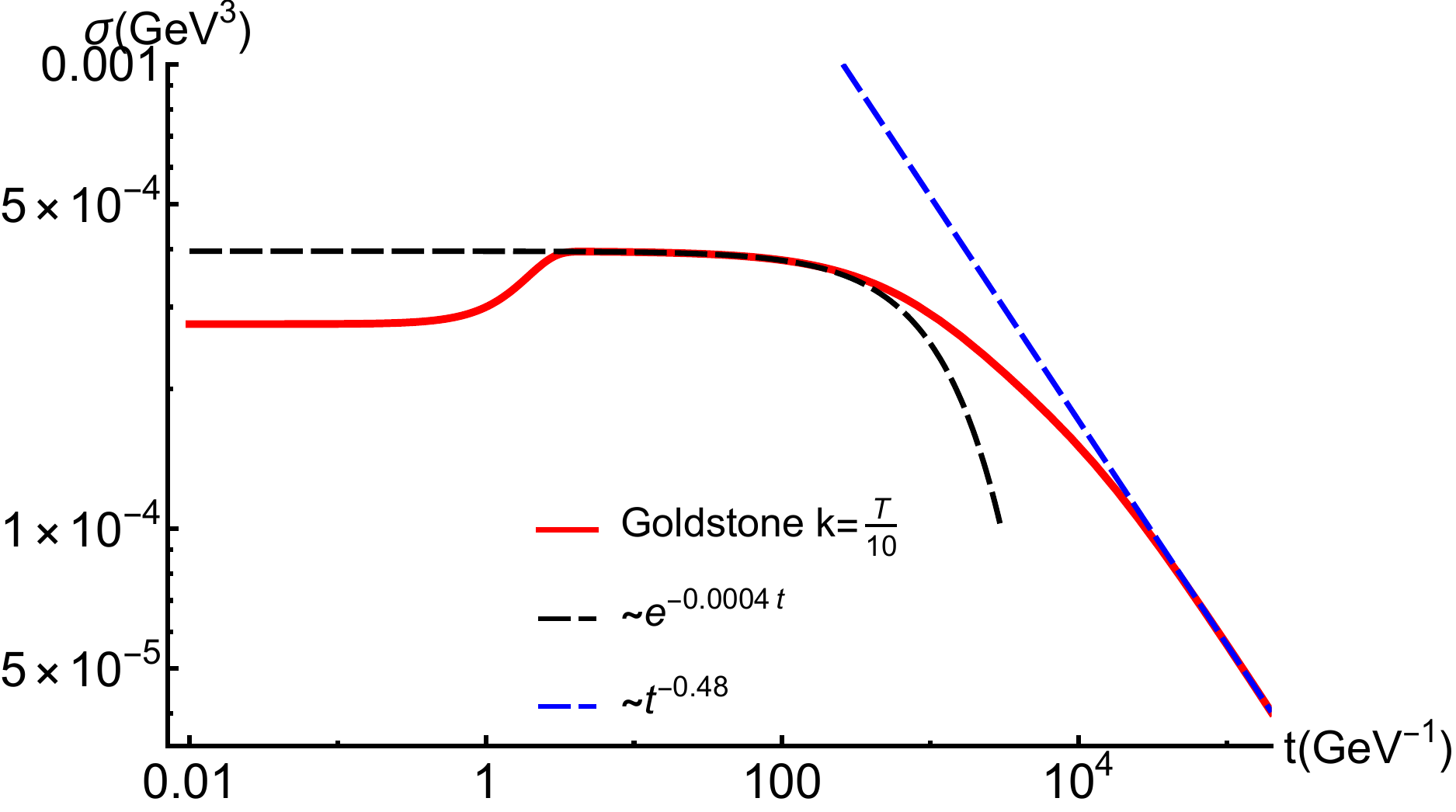}
    \put(20,10){\bf{(b)}}
    \end{overpic}
    \caption{Time evolution curves of chiral condensate for the systems (a) without introducing Goldstone modes, and (b) involving Goldstone mode with spatial momentum $k=T/10$. The black dashed lines depict the exponential fit $e^{-\Gamma t}$ of the approximate plateau stage. Note the fitted values of $\Gamma$ are of order $\mathcal{O}(10^{-4})$, implying some kinds of "equilibrium" have been established in such stage. The Blue dashed lines describe the power law behaviors of final relaxation stage.}
    \label{fig:criticalChiralFit}
\end{figure}

Under careful investigation, the dynamical evolution can be more precisely categorized into three different stages after the onset of prethermalization. The first stage is well described by exponent function $\sim e^{-\Gamma t}$ with rather small value of $\Gamma$. The long-time stage is related to the relaxation to final thermal states and can be fitted into some power law behaviors $\sim t^{\alpha}$ with $\alpha<0$. The intermediate stage with time $t$ approximately ranging from $10^3 {\rm{GeV}}^{-1}$ to $10^4 {\rm{GeV}}^{-1}$ behaves like neither simple exponent nor some specific power law behaviors. To clarify, we fit the evolution curves by using simple exponent function and power law function and the numerical results are displayed in Fig. \ref{fig:criticalChiralFit}. One can see that the fitted values of $\Gamma$ are of typical order $\mathcal{O}(10^{-4})$ which approximately can be treated as zero. This suggests that some kinds of "equilibrium" have been established in such stage. Consequently, we identify such stage as "prethermalization" naturally.      

\begin{figure}[htbp]
    \centering
    \begin{overpic}[width=0.45\textwidth]{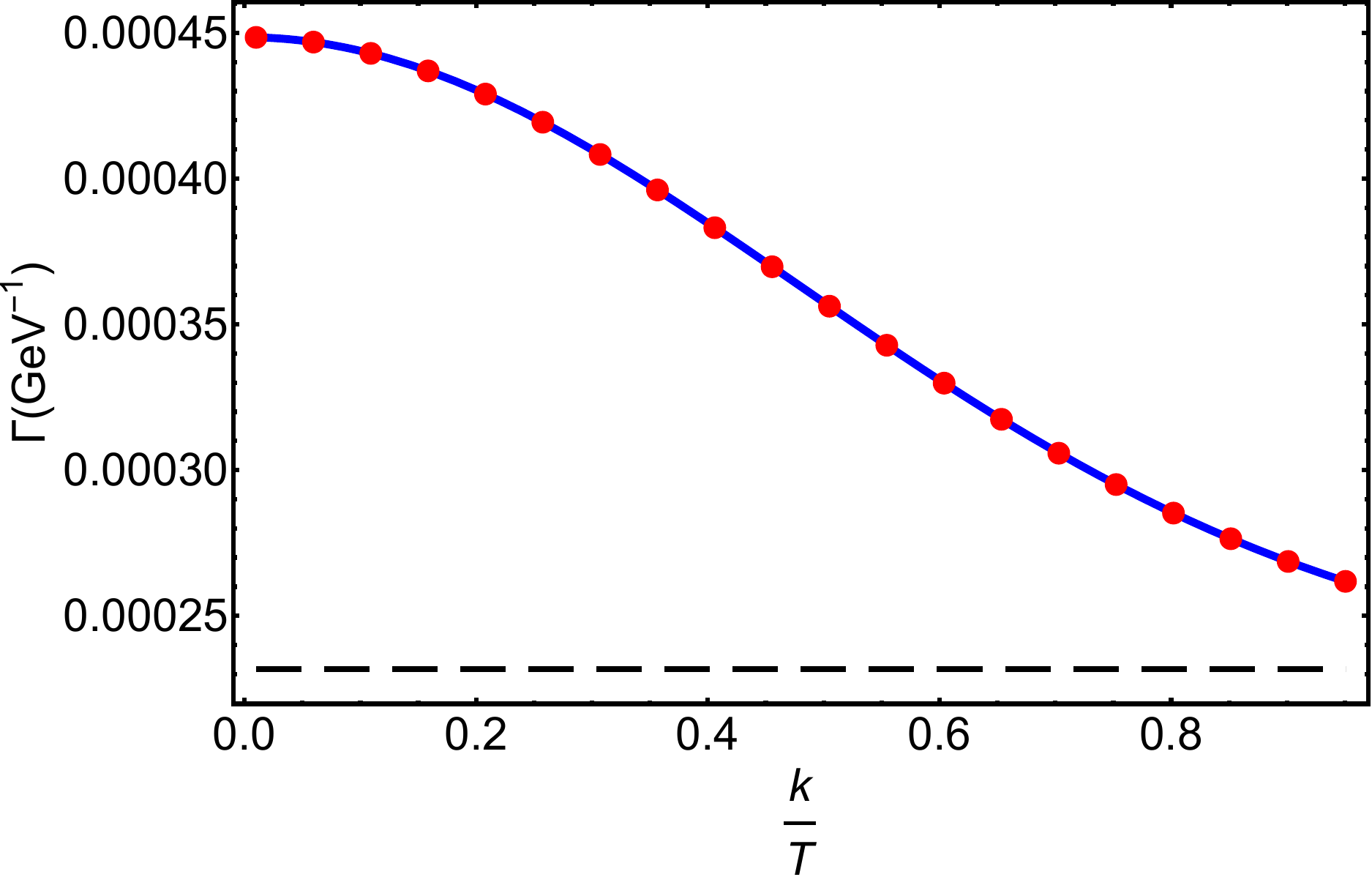}
    \put(90,50){\bf{(a)}}
    \end{overpic}
        \begin{overpic}[width=0.45\textwidth]{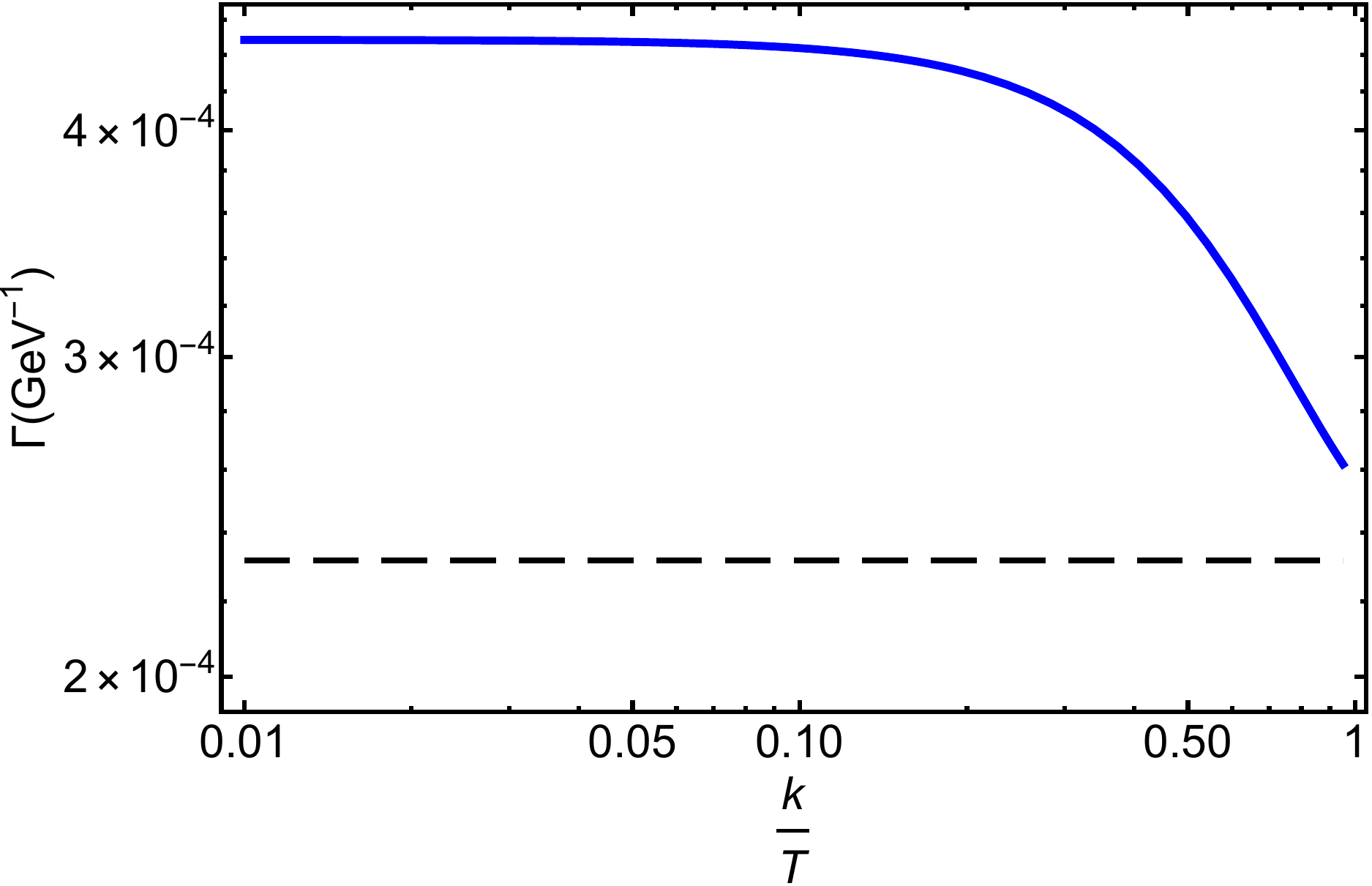}
    \put(90,50){\bf{(b)}}
    \end{overpic}
    \caption{(a) The Goldstone spatial momentum dependence of decay rate $\Gamma$. (b) Double-logarithmic plot of the same relation as (a). The black dashed lines in two figures show the decay rate without introducing Goldstone modes and the red circles refer to the original calculated data. One observes immediately that the large momentum case should approach to the no Goldstone limit. This because Goldstone modes with large spatial momentum decay to zero quickly and decouple from other degrees of freedom.}
    \label{fig:prethermalDecay}
\end{figure}

Despite the rather small value of $\Gamma$, this quantity shows non-trivial Goldstone spatial momentum dependence. This is demonstrated in Fig. \ref{fig:prethermalDecay}. In general, with the increasing Goldstone spatial momentum, the decay rate $\Gamma$ will gradually approach to the limit without Goldstone mode. In fact, the Goldstone modes with large spatial momentum quickly decay to zero and decouple from the whole system. On the contrary, soft Goldstone modes have apparent effects on the decay behavior in the prethermalization stage. Note in the zero momentum limit, the decay rate $\Gamma$ is approximately twice larger than the one obtained without involving Goldstone modes. This implies introducing Goldstone modes in the system could possibly break the transient stationary states established in the prethermalization stage and make the system evolve into final thermal states much earlier.  

\begin{figure}[ht]
\centering
\includegraphics[scale=0.45]{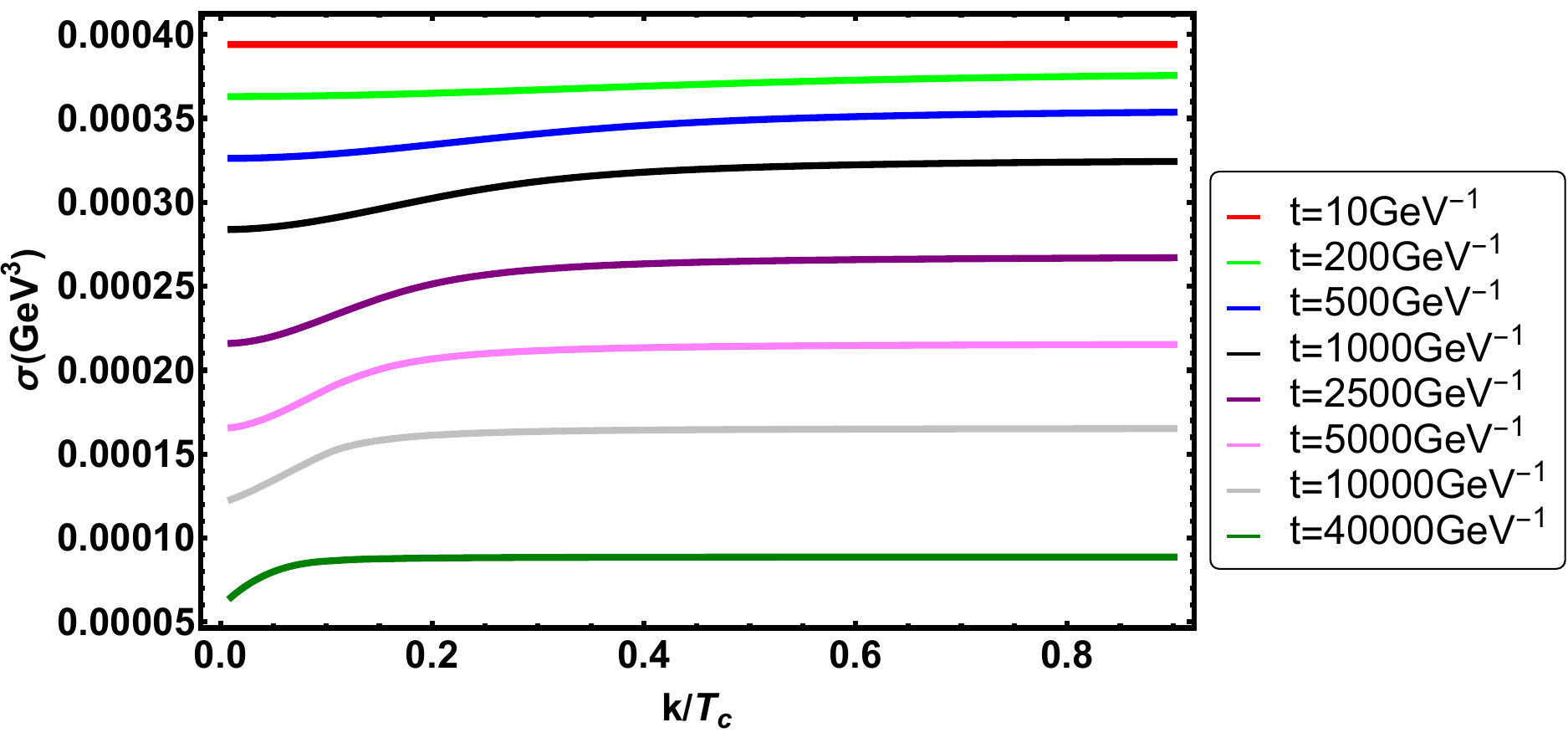}
 \caption{The momentum distribution of chiral condensate at different times. We observe the non-trivial momentum dependence appearing only with soft momentum which are approximately in the range of $k/T_c\lesssim 0.1$.}
   \label{fig:CriMomDis}
\end{figure}

Remarkably, we also observe that the decay rate $\Gamma$ is almost independent of spatial momentum for $k/T_c\lesssim 0.1$. This could be interpreted as universality to some extent. Interestingly, we note that this particular momentum range overlaps with the range observed in Fig. \ref{fig:CriMomDis}. It is demonstrated in Fig. \ref{fig:CriMomDis} that the non-trivial momentum dependence in the dynamical evolution also arises in similar range. We believe there remains some connection between these two phenomena, but the clear illustration about this point is still missing. 

\begin{figure}[htbp]
    \centering
    \begin{overpic}[width=0.70\textwidth]{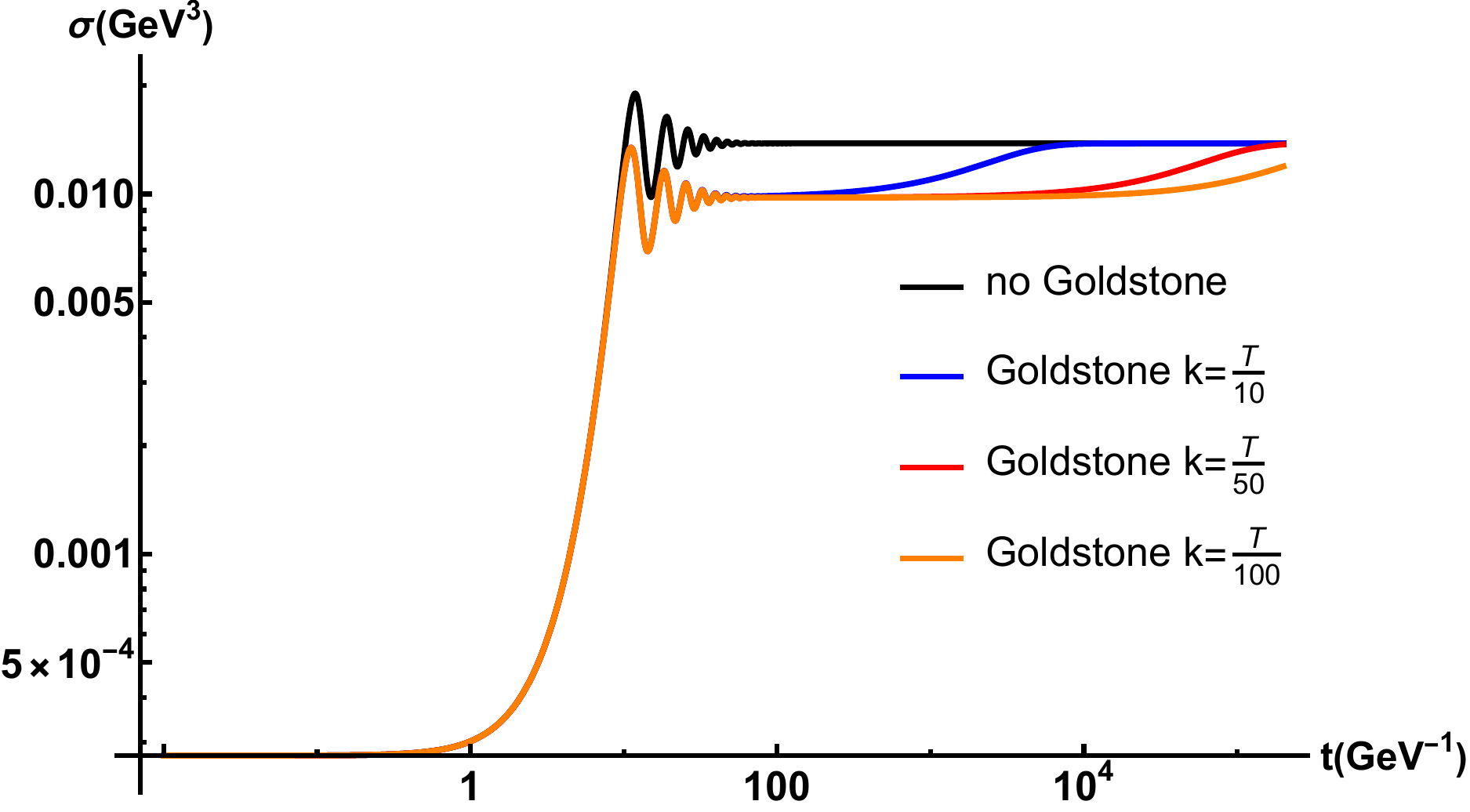}
    \put(20,50){\bf{(a)}}
    \end{overpic}
        \begin{overpic}[width=0.70\textwidth]{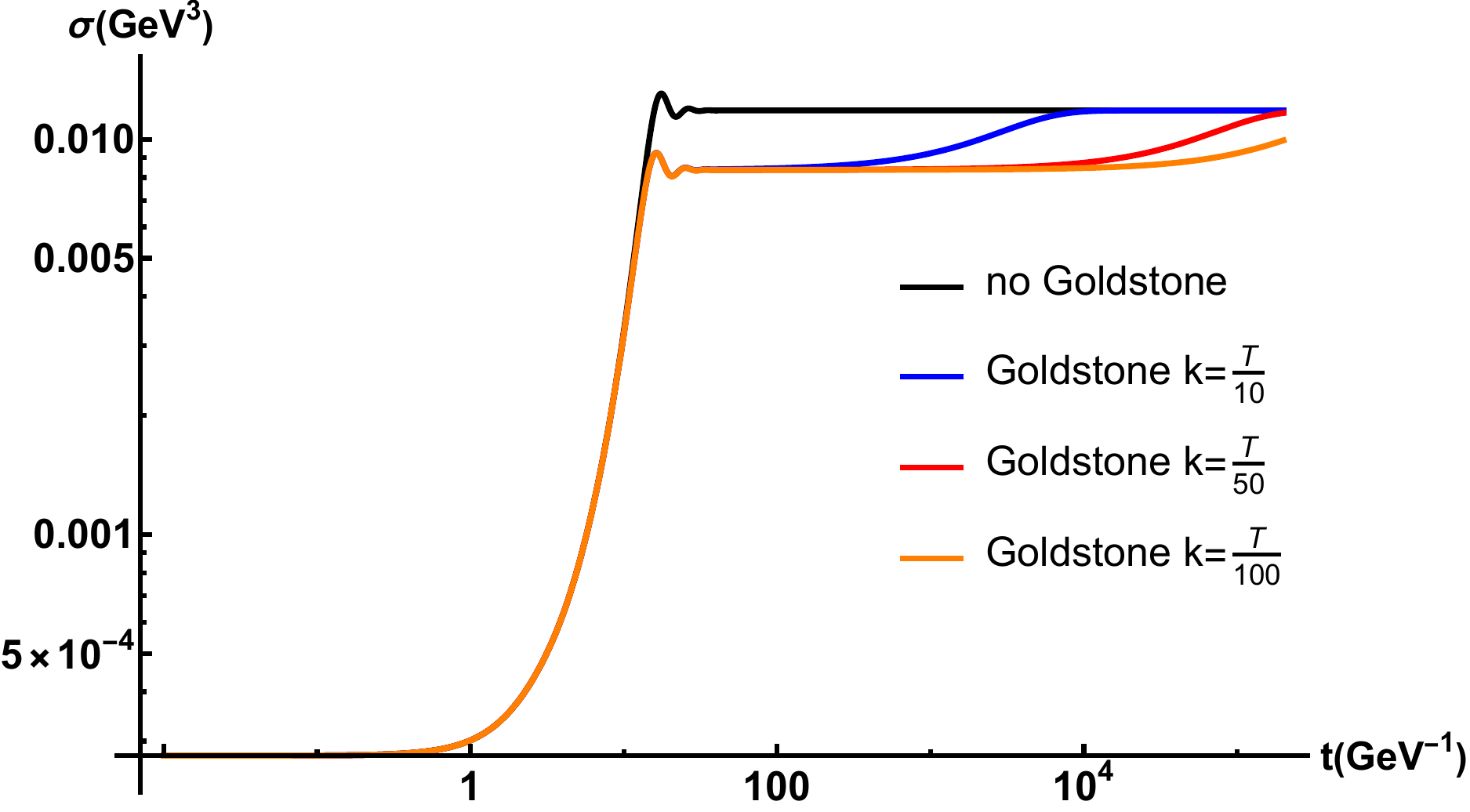}
    \put(20,50){\bf{(b)}}
    \end{overpic}
    \caption{The typical evolution curves of chiral condensate at temperature $T<T_c=163{\rm{MeV}}$. Two cases with different temperatures (a) $T=120{\rm{MeV}}$ and (b) $T=140{\rm{MeV}}$ have been shown here. At each given temperature, we compare the dynamical evolution behavior of chiral condensate without introducing any Goldstone modes in the system to those cases involving Goldstone modes with different values of spatial momentum. Apparent intermediate stage appears in the dynamical evolution when coupling Goldstone modes to the system. We refer this stage to prethermalization at non-critical temperature region.  }
    \label{fig:chiralMomCompare}
\end{figure}

\subsubsection{Numerical result at non-critical temperature}
\label{subsubsec:noncritical temperature}
Unlike the critical case, the most striking phenomenon involving Goldstone modes appears at non-critical temperature $T<T_c$. Above $T_c$, the Goldstone modes will decay rather rapidly and decouple from the whole system, which is just as we expected.
 
In the previous study \cite{Cao:2022mep,Flory:2022uzp}, the possible prethermalization seems to be appearing only at critical temperature, or at least very close to the critical temperature. When introducing extra degrees of freedom which are related to the Goldstone modes, we find that the prethermalization stage can be present in the intermediate time scale of dynamical evolution even at temperature far away from critical point. In Fig. \ref{fig:chiralMomCompare}, we choose two temperature $T=120{\rm{MeV}}$ and $T=140{\rm{MeV}}$ as representatives. From the direct numerical study, we clearly observe the appearance of intermediate stage in the whole dynamical evolution when introducing some Goldstone modes in the system, which is very different from the typical behavior of the case with only order parameter. Similar to the intermediate stage appearing in the critical temperature case, we designate this stage as prethermalization occurring at a non-critical temperature. 

Note the duration of this intermediate stage increases drastically, ranging from several time scales, when the spatial momentum of Goldstone modes decreases. Accordingly, we conjecture that the final relaxation time which is related to the long-time thermalization stage will approach to infinity when the spatial momentum of Goldstone $k$ takes its limit value zero. Physically, considering the Goldstone nature of such mode, this limit can not be achieved exactly. In practical numerical calculation, the major obstacle comes from the large numerical instability in this zero momentum limit. Thus, we only treat above statement as some kind of conjecture. This very slow relaxation behavior is similar to the critical slowing-down which typically arises at the true thermal critical point. In essence, the critical point should correspond to some fixed point in physical parameter space. This suggests that the zero momentum limit could possibly correspond to some non-thermal fixed point hidden in the non-equilibrium dynamics.    

\begin{figure}[ht]
    \centering
    \includegraphics[scale=0.4]{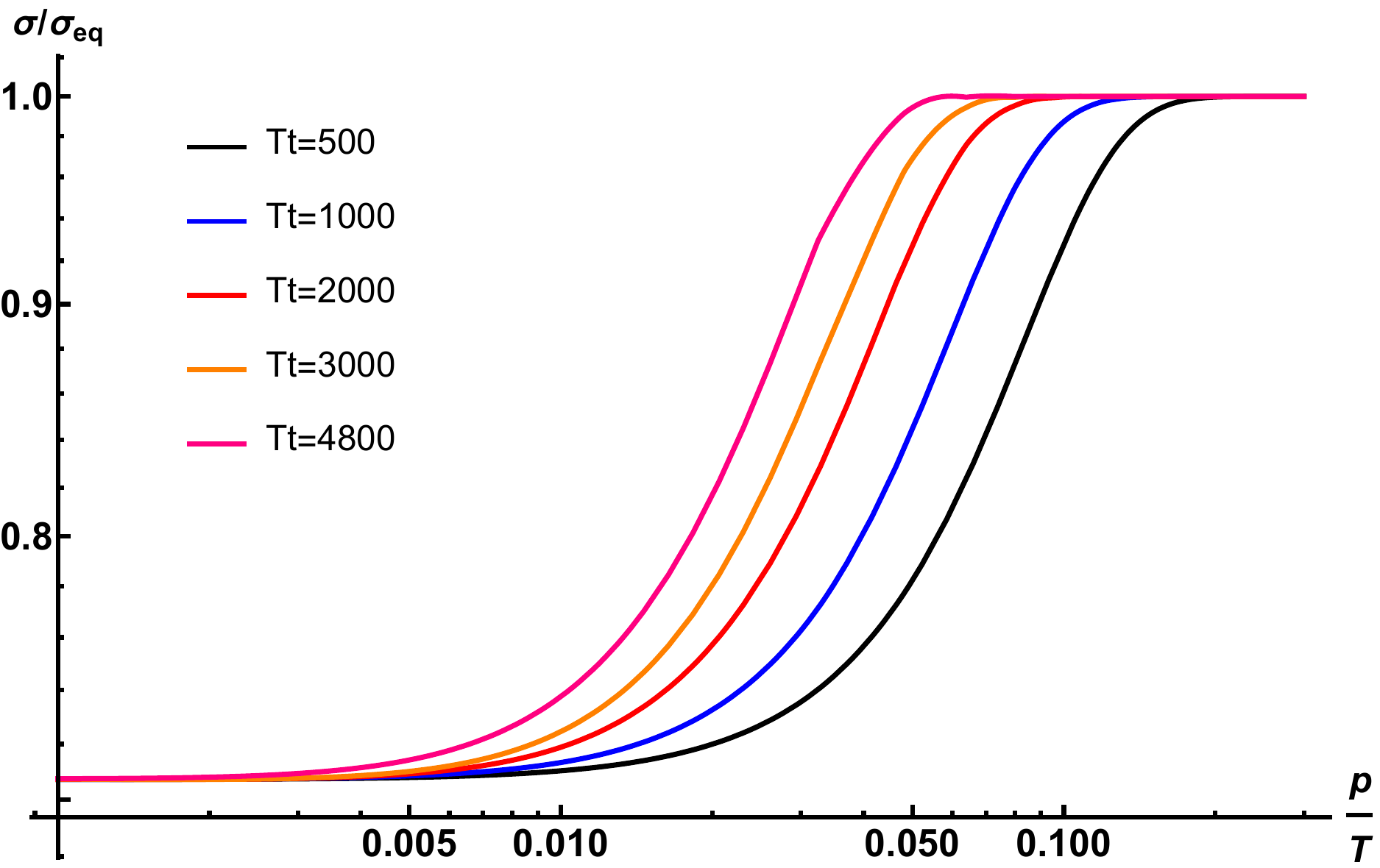}
   \caption{The momentum distribution of chiral condensate with $T=130\rm{MeV}$ at different times. Here, $T$ is identified as the temperature of final equilibrium state and $t$ is just physical time. The different values of corresponding dimensionless combination $Tt$ have been used to represent different moments during time evolution. We observe the non-trivial momentum dependence appearing only in some narrow momentum region. }
   \label{fig:MomDis}
\end{figure}

Obviously, the momentum dependence of non-equilibrium dynamics needs to be studied. The momentum distribution of chiral condensate at different times is presented in Fig. \ref{fig:MomDis}. In order to make comparison clear, we rescale chiral condensate with its thermal equilibrium value $\sigma_{eq}$ at given temperature. Most remarkably, we observe clear self-similar evolution behavior. 

\begin{figure}[htbp]
    \centering
    \begin{overpic}[width=0.45\textwidth]{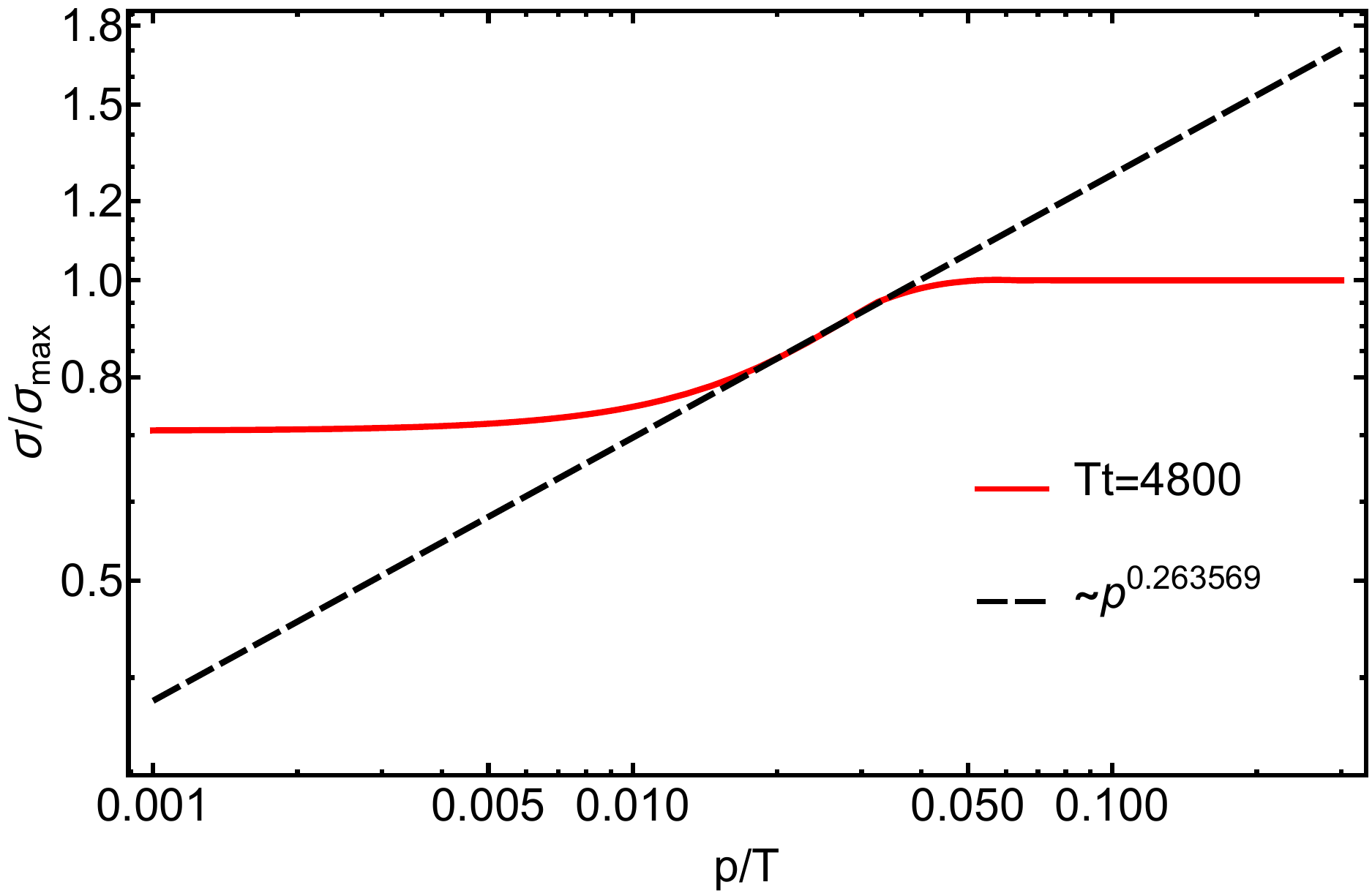}
    \put(15,50){\bf{(a) $T=120\rm{MeV}$}}
    \end{overpic}
    \begin{overpic}[width=0.45\textwidth]{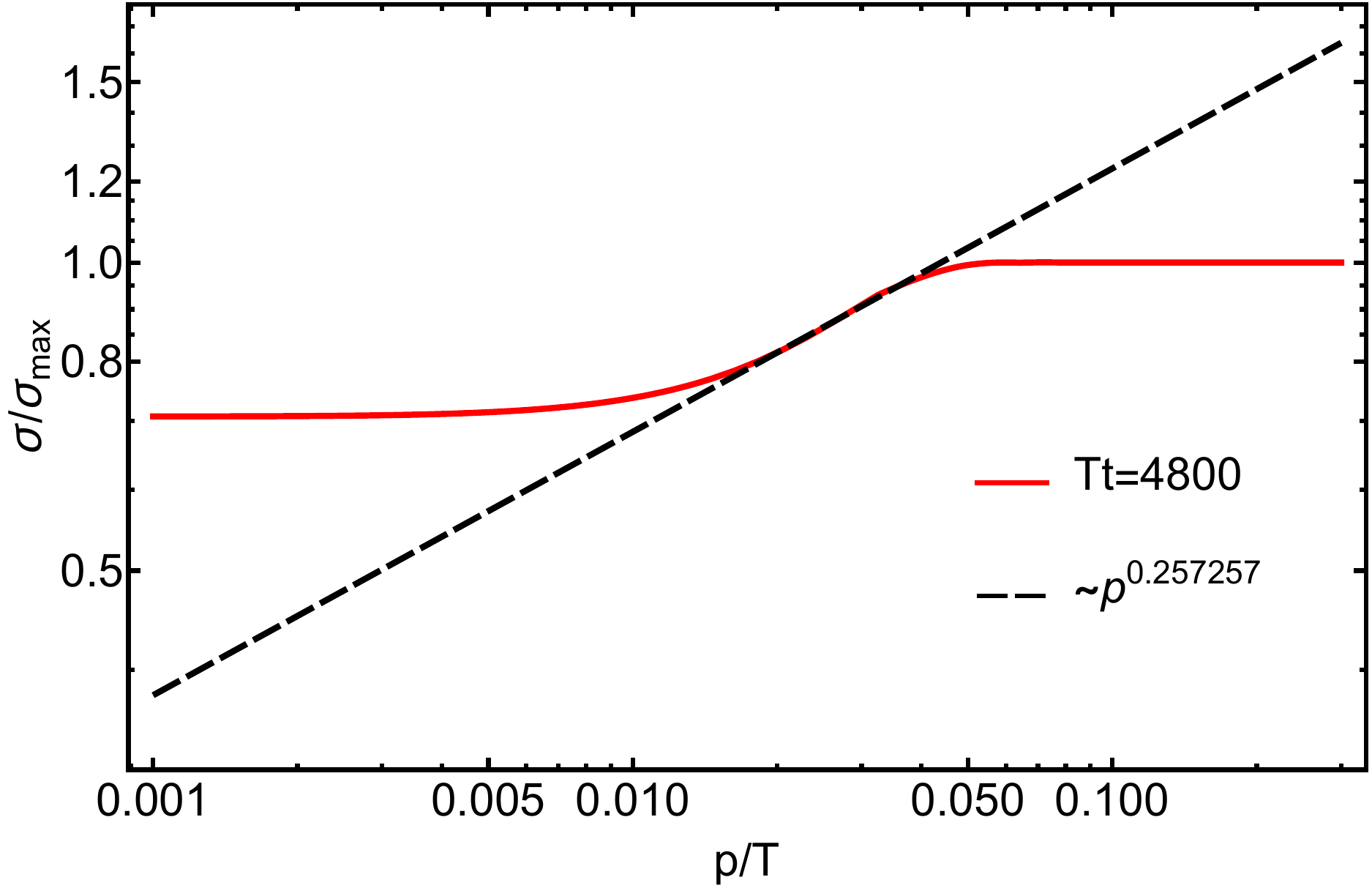}
    \put(15,50){\bf{(b) $T=130\rm{MeV}$}}
    \end{overpic}
    \begin{overpic}[width=0.45\textwidth]{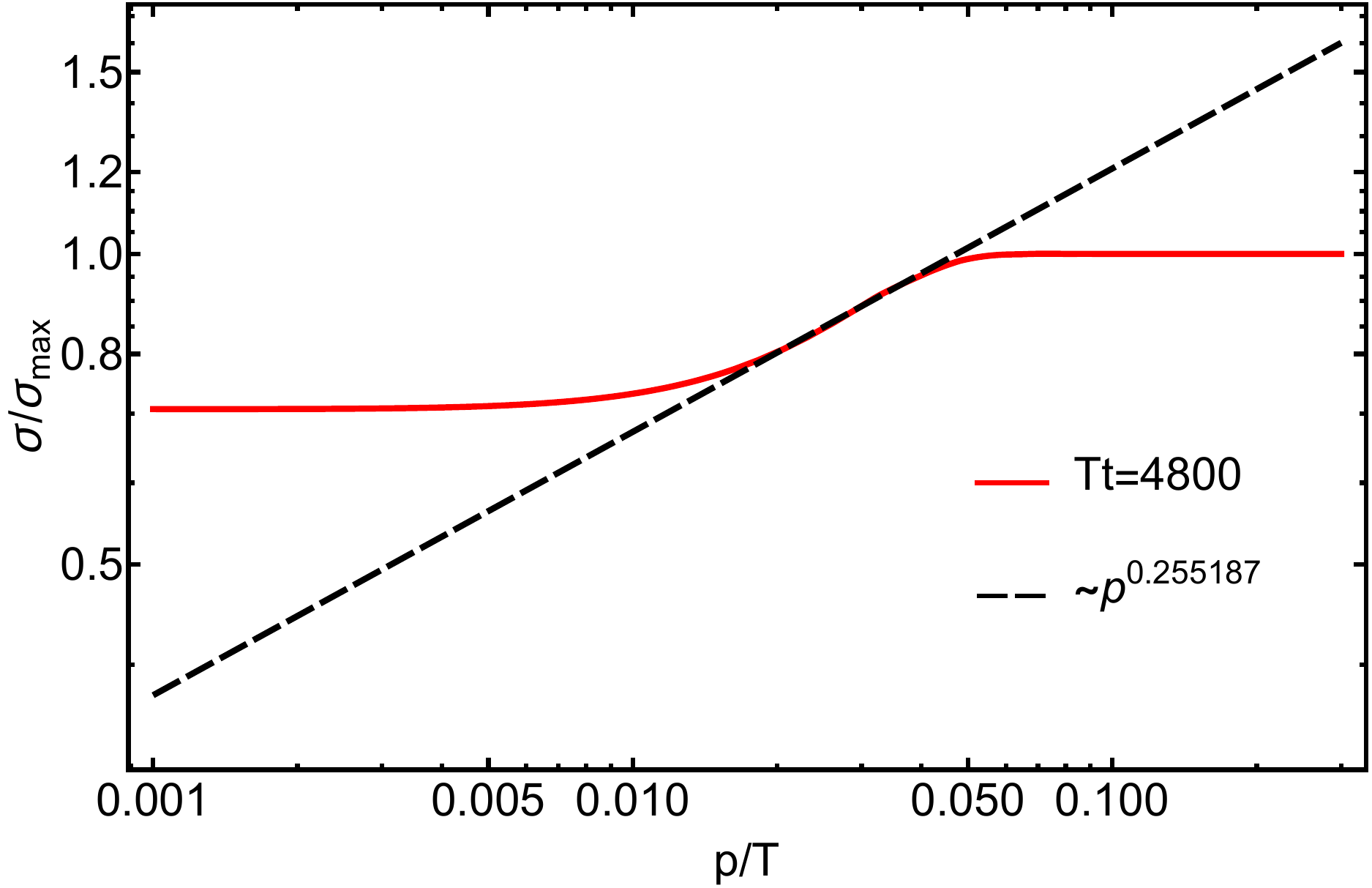}
    \put(15,50){\bf{(c) $T=140\rm{MeV}$}}
    \end{overpic}
    \begin{overpic}[width=0.45\textwidth]{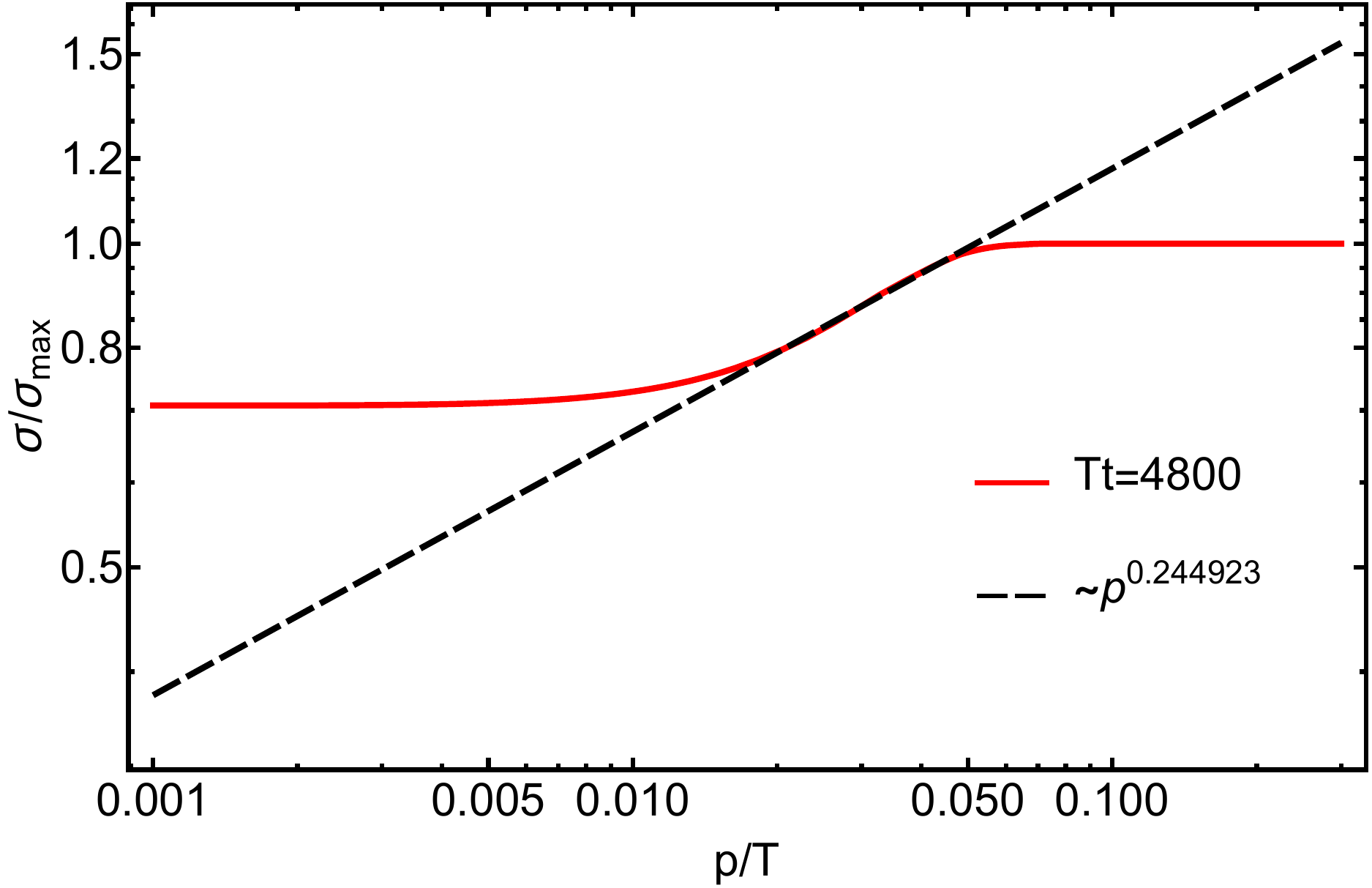}
    \put(15,50){\bf{(d) $T=150\rm{MeV}$}}
    \end{overpic}
    \caption{The power law behavior of momentum dependence of chiral condensate with (a) $T=120\rm{MeV}$, (b) $T=130\rm{MeV}$, (c) $T=140\rm{MeV}$ and (d) $T=150\rm{MeV}$ at $Tt=4800$. Here, $\sigma_{\rm{max}}$ corresponds to the maximal chiral condensate value at certain time. Typically, this value equals to the chiral condensate of thermal equilibrium state at given temperature. The power laws with approximate exponent $\alpha=1/4$ show little temperature dependence.}
    \label{fig:MomFit}
\end{figure}

Note that the system coupling to Goldstone modes with typical spatial momentum $p$ in the range $p/T\gtrsim 0.1$, which we will refer to hard region, approaches to thermal equilibrium very quickly. In Fig. \ref{fig:MomDis}, even at $Tt=500$, which corresponds to $t=3846\rm{GeV^{-1}}$, the system in hard region has already been in thermal equilibrium. The non-trivial momentum dependence only manifests itself only in a rather narrow band. We also observe that this narrow band translates to the softer momentum region when the system continues to evolve. This observation agrees with our physical intuition. The Goldstone modes with large spatial momentum will quickly decay to zero and as a consequence, decouple from the whole system. For this reason, the Goldstone modes with soft spatial momentum dominate in the non-equilibrium dynamics.

The non-trivial momentum dependence can be described by some approximate power law $\sigma/\sigma_{eq}\sim (p/T)^{\alpha}$. In Fig. \ref{fig:MomDis}, we demonstrate that the exponent $\alpha$ has some time dependence. However, direct calculation shows that temperature change has very little effects on this particular exponent $\alpha$. Indeed, We see in Fig. \ref{fig:MomFit} that the fitted values of $\alpha$ at four different final temperatures are very close to $1/4$. Regarding the possible numerical error in fit procedure, we conclude that the typical momentum dependence of chiral condensate in non-equilibrium evolution show some universality with respect to temperature.        

In general, the evolution in the prethermalization stage is expected to depend on the information of initial state, at least partially. So the study of the initial condition dependence is required. Note that there are two major initial parameters $\sigma_{\rm{ini}}$ and $\pi_{\rm{ini}}$ in the system. The relative ratio of these two parameters also play an important role during the dynamical evolution. To clarify, we first consider the cases with different values of $\sigma_{\rm{ini}}$ but fixed $\sigma_{\rm{ini}}/\pi_{\rm{ini}}$ ratio.       
\begin{figure}[htbp]
    \centering
    \begin{overpic}[width=0.70\textwidth]{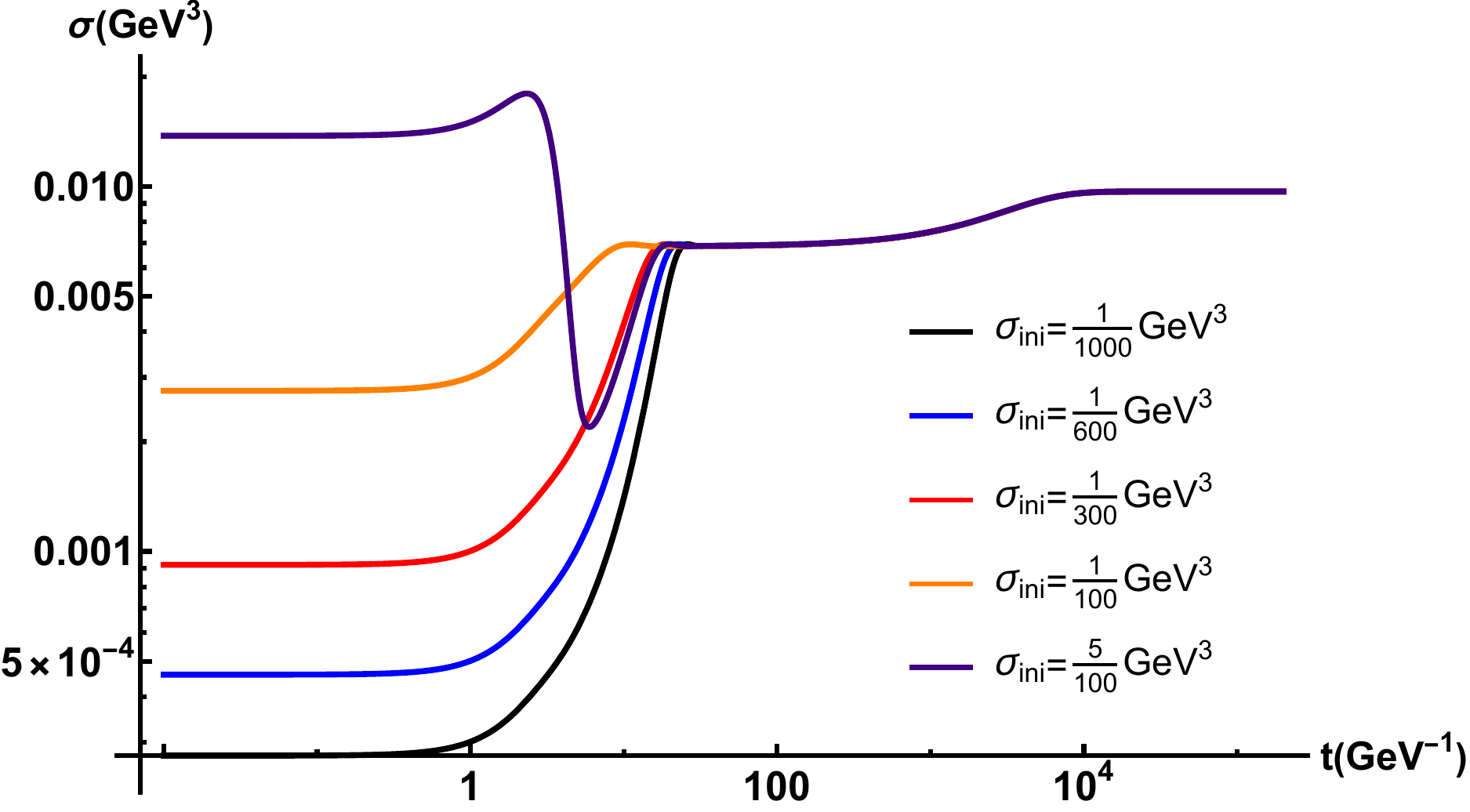}
    \put(90,50){\bf{(a)}}
    \end{overpic}
        \begin{overpic}[width=0.70\textwidth]{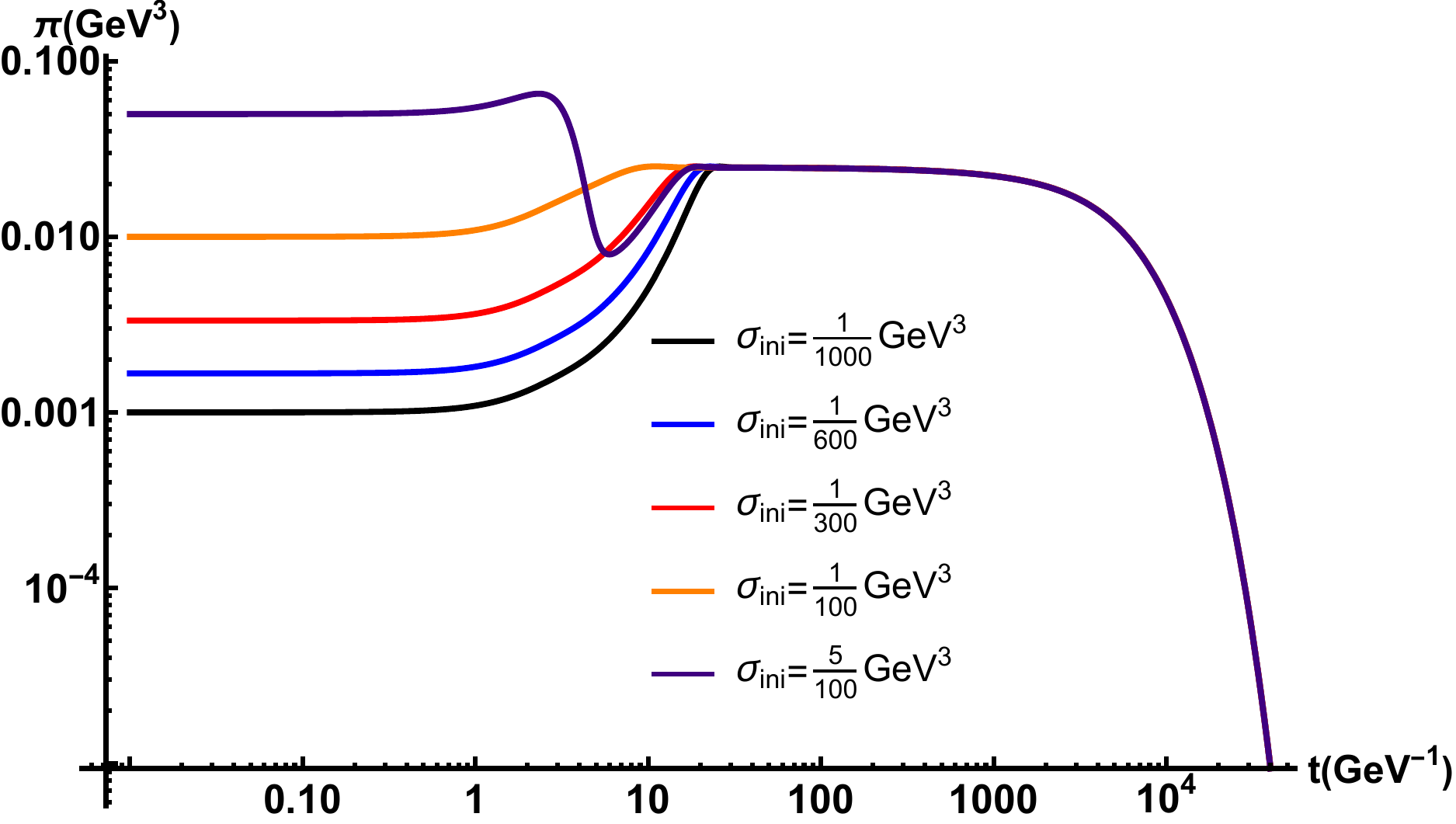}
    \put(90,50){\bf{(b)}}
    \end{overpic}
    \caption{Comparison of time evolution behavior with different initial states, fixed background temperature $T=160{\rm{MeV}}$ and fixed spatial momentum of Goldstone mode $k=T/10$. Besides, the ratio $\sigma_{\rm{ini}}/\pi_{\rm{ini}}=1$ is also fixed. (a) Evolution curves of chiral condensate with different $\sigma_{\rm{ini}}$. (b) Evolution curves of Goldstone mode with different $\sigma_{\rm{ini}}$.}
    \label{fig:iniRationFixed}
\end{figure}

In Fig. \ref{fig:iniRationFixed}, we choose $\sigma_{\rm{ini}}/\pi_{\rm{ini}}=1$ and background temperature $T=160{\rm{MeV}}$. To make the effect of Goldstone mode more explicit, we also set its spatial momentum $k=T/10=16{\rm{MeV}}$. Five different values of $\sigma_{\rm{ini}}$ have been chosen. Although the precise short-time dynamics are dramatically different, the evolution behavior of two physical quantities in the intermediate prethermalization stage and long-time thermalization stage are exactly the same. From the physical intuition, since the system lose all the information about initial states in the final long-time evolution, the independence of initial states in this stage is easy to understand. However, the universal behavior emerging from the prethermalization stage is beyond naive expectation. Actually, the prethermalization stage should involve partial information about initial states, otherwise the system is in fact in thermal equilibrium state. Accordingly, the evolution in the intermediate time scale can not be totally independent of initial conditions and partial information should be maintained. 

\begin{figure}[htbp]
    \centering
    \begin{overpic}[width=0.70\textwidth]{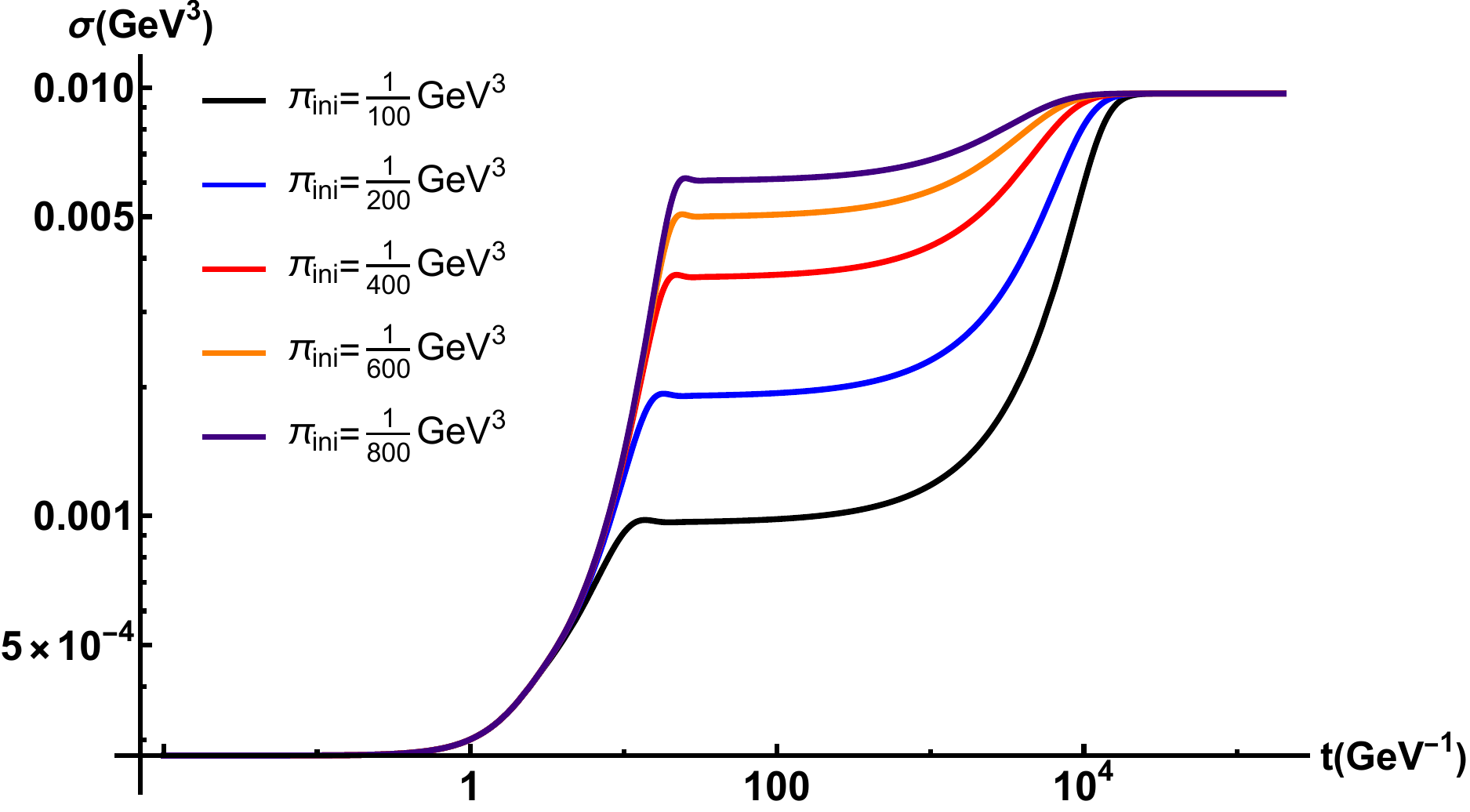}
    \put(90,45){\bf{(a)}}
    \end{overpic}
        \begin{overpic}[width=0.70\textwidth]{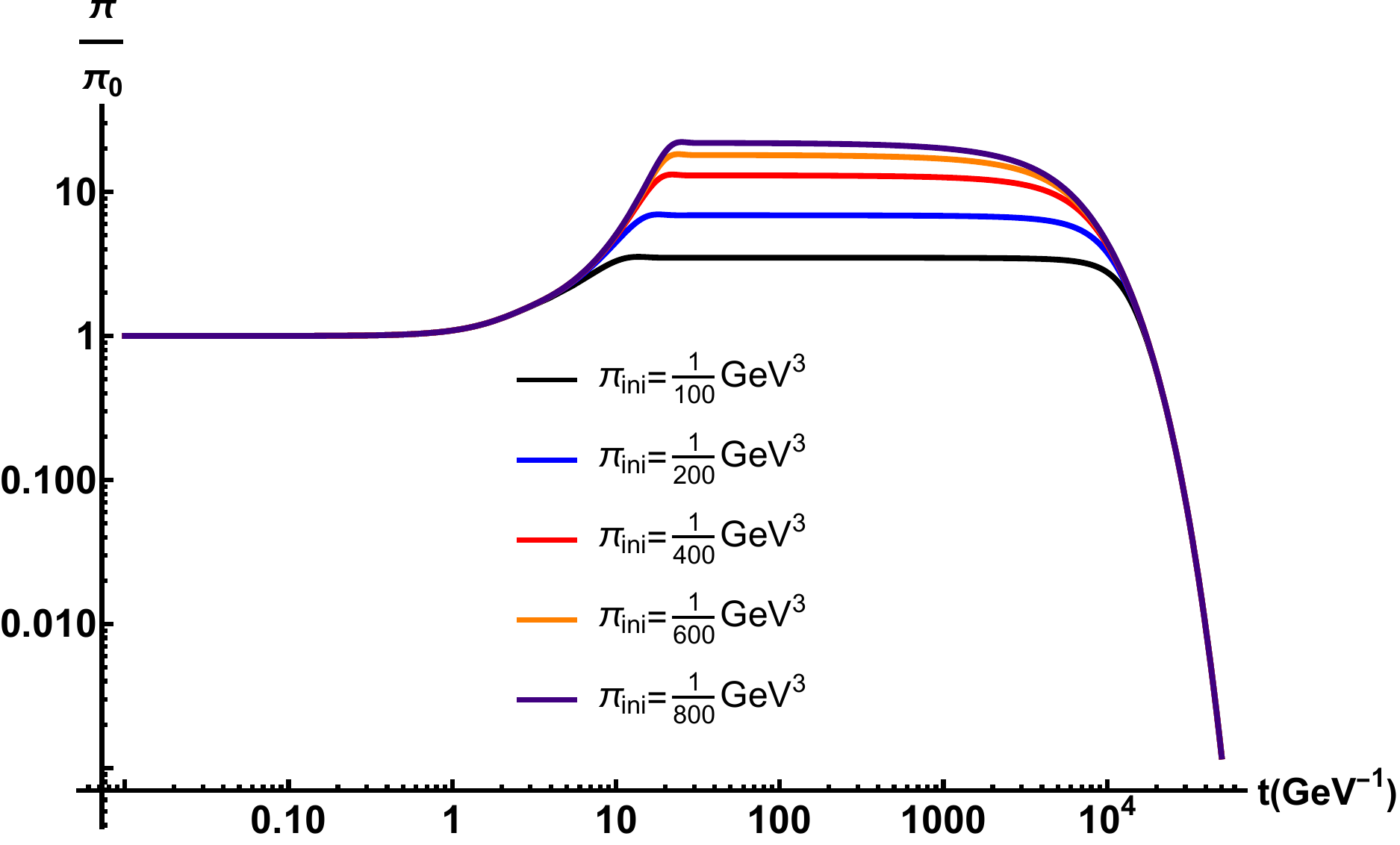}
    \put(90,45){\bf{(b)}}
    \end{overpic}
    \caption{Comparison of time evolution behavior with different $\pi_{\rm{ini}}$, fixed $\sigma_{\rm{ini}}=1/1000{\rm{GeV}}^3$ and fixed spatial momentum of Goldstone mode $k=T/10$. (a) Evolution curves of chiral condensate. (b) Evolution curves of Goldstone mode. In order to compare the intermediate stage, we rescale Goldstone mode $\pi$ with its initial value $\pi_0$.}
    \label{fig:iniRationChange}
\end{figure}

The desired initial value dependence of the prethermalization is encoded in the ratio $\sigma_{\rm{ini}}/\pi_{\rm{ini}}$. In Fig. \ref{fig:iniRationChange}, we directly compare evolution of physical quantities with different $\sigma_{\rm{ini}}/\pi_{\rm{ini}}$ ratio. Clearly, when changing the ratio, the actual values of $\sigma$ and $\pi$ in the prethermalization stage and the relaxation rate to thermal equilibrium can be modified drastically. The larger $\pi_{\rm{ini}}$ will lead to smaller $\sigma$ value in the intermediate stage and larger relaxation rate. But the duration of prethermalization do not change a lot. In some wider sense, this kind of behavior shows universality behind non-equilibrium evolution. Unfortunately, this possible universality, especially its quantitative properties, remains unclear and needs to be further studied.   

\begin{figure}[htbp]
    \centering
    \begin{overpic}[width=0.70\textwidth]{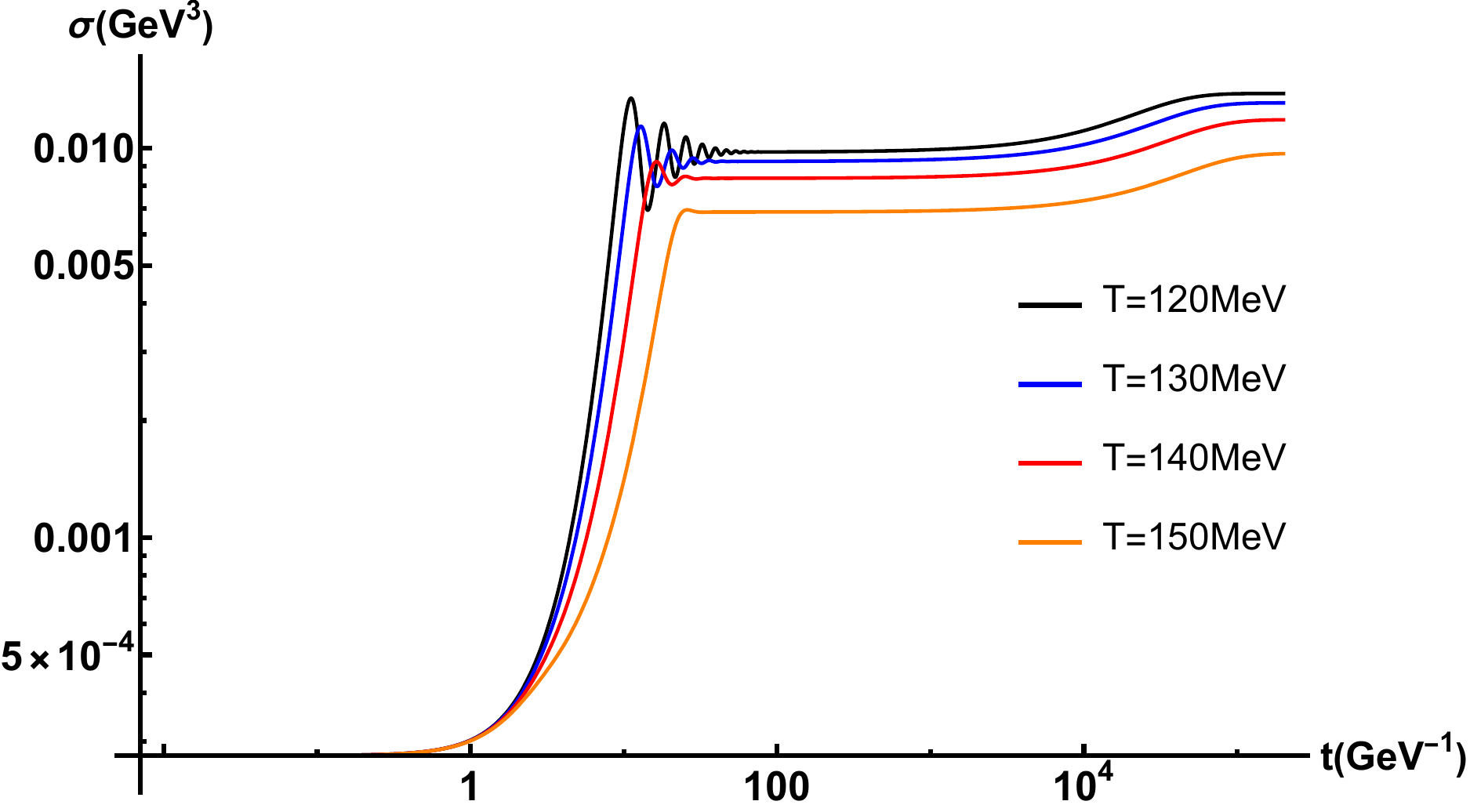}
    \put(20,20){\bf{(a)}}
    \end{overpic}
        \begin{overpic}[width=0.70\textwidth]{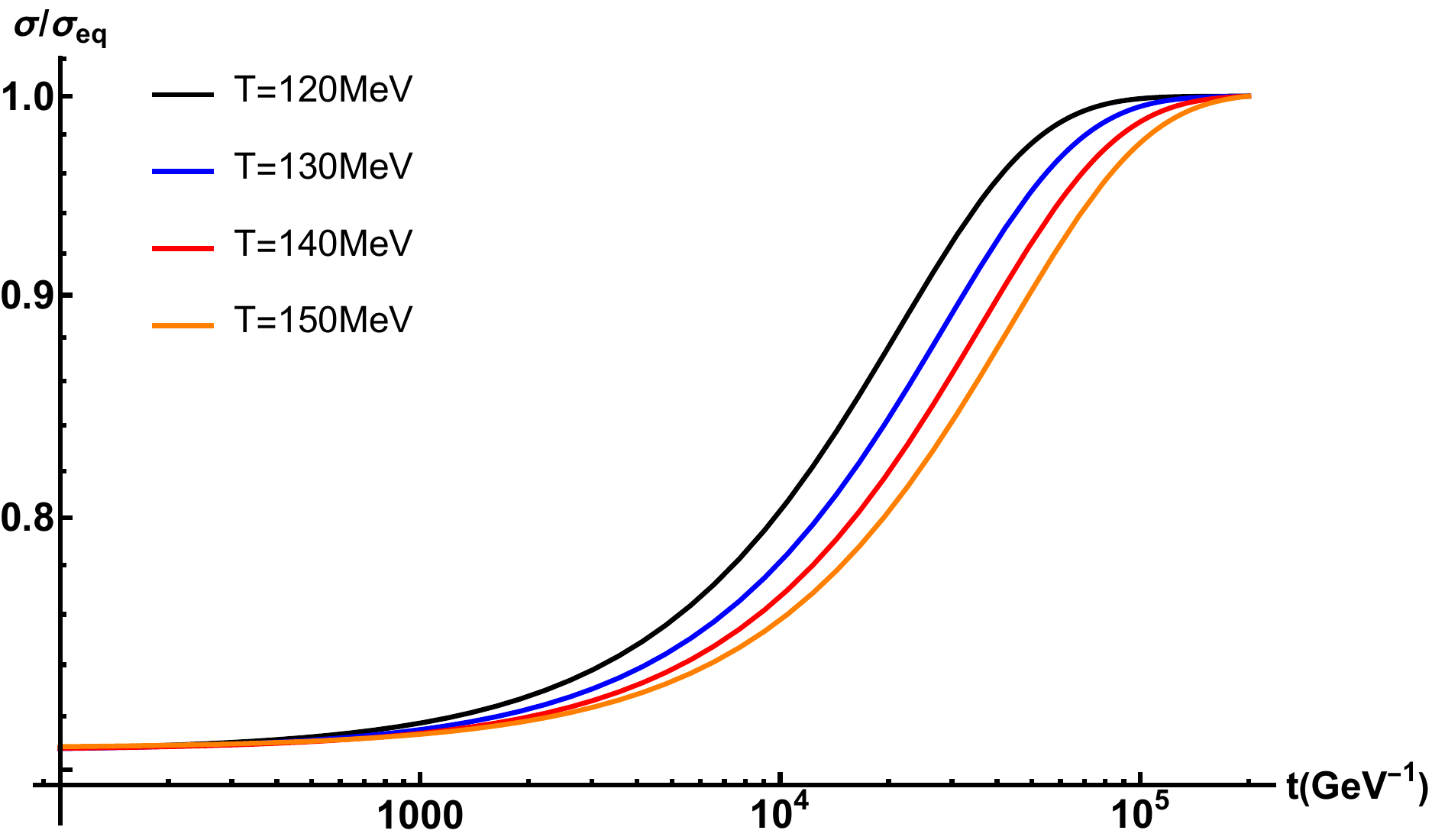}
    \put(20,20){\bf{(b)}}
    \end{overpic}
    \caption{Comparison of time evolution behavior at different temperatures with fixed $\sigma_{\rm{ini}}=1/1000{\rm{GeV}}^3$, initial ratio $\sigma_{\rm{ini}}/\pi_{\rm{ini}}=1$ and spatial momentum of Goldstone mode $k=4\rm{MeV}$. (a) Evolution curves of chiral condensate. (b) Evolution curves of $\sigma/\sigma_{\rm{eq}}$, with time $t$ ranging from $100{\rm{GeV}}^{-1}$ to $2\times 10^5{\rm{GeV}}^{-1}$, which covers entire prethermalization and relaxation parts.}
    \label{fig:temCompare}
\end{figure}

Although we can not fully understand all the physical properties related to prethermalization, we indeed observe some interesting universal behaviors of this particular dynamical stage. Apart from parameters discussed before, we also study whether the temperature of final equilibrium state could have some apparent effects on the entire non-equilibrium evolution. This is demonstrated in Fig. \ref{fig:temCompare}. Typically, the absolute values of chiral condensate in the intermediate stage are different from each other. Nevertheless, we observe that the rescaled chiral condensates $\sigma/\sigma_{\rm{eq}}$ approach the same numerical value in the intermediate stage of dynamical evolution. This suggests that the rescaled chiral condensates in the prethermalization stage at different temperatures do not rely on the information of final thermal states and show the desired universal behavior. On the other hand, there remains some subtleties in this discussion. To clarify, note that the relaxation towards different final thermal states start at different times. Consequently, the full dynamical evolution is actually dependent on final states, at least partially. This seems a weird observation of non-equilibrium dynamics.            

\subsection{Universal scaling behaviors at fixed points}
\label{subsec:thermal}
In the following, we investigate the universal scaling behaviors related to different kinds of fixed point. A well-known fact about critical point of second order phase transition is that this point can be understood as fixed point in physical parameter space. There are certain universal scaling relations related to such thermal fixed point. Besides, interesting universal behavior manifests itself during the non-equilibrium dynamical evolution. This particular unexpected scaling relation is conjectured to be related to non-thermal fixed point.

\subsubsection{Critical slowing down}
\label{subsubsec:criticalSD}
The typical phenomenon related to the relaxation process at critical temperature is the so-called critical slowing down. In the thermodynamic limit, the correlation length is divergent at exact critical point and the related behavior can be understood by universal scaling relations. The derivation in this part closely follow previous work \cite{Cao:2022mep}.

In this part, the only scaling relation we are interested in can be simply written as
\begin{align}
    \label{eq:3.3}
    \xi(t)=b\xi(tb^{-z}),
\end{align}
with the dynamical critical exponent $z$ and rescaling factor $b$. Here, $\xi(t)$ refers to time-dependent correlation length. From eq. \ref{eq:3.3}, it is suggested that the order parameter should satisfy
\begin{align}
   \label{eq:3.4}
   \sigma(t)=b^{-\beta/\nu}\sigma(tb^{-z}),
\end{align}
with standard static critical exponents $\beta$ and $\nu$. In this case, one can choose the value of rescaling factor $b$ such that $tb^{-z}\sim {\rm{constant}}$. Thus, from eq. \ref{eq:3.4}, the scaling form of order parameter in terms of time $t$ can be obtained as
\begin{align}
    \label{eq:3.5}
    \sigma(t)=t^{-\beta/\nu z}f({\rm{number}})\propto t^{-\beta/\nu z}.
\end{align}
This relation directly shows that infinite time is needed to achieve the final thermal state with $\sigma_{\rm{eq}}=0$. This phenomenon is exactly the famous critical slowing down. In our soft-wall model, the mean-field static critical exponents have been applied (order parameter critical exponent $\beta=1/2$ and correlation length critical exponent $\nu=1/2$). Consequently, the scaling form of order parameter can be simplified as
\begin{align}
    \label{eq:3.6}
    \sigma(t)\propto t^{-1/z}.
\end{align}

\begin{figure}[ht]
    \centering
    \includegraphics[scale=0.4]{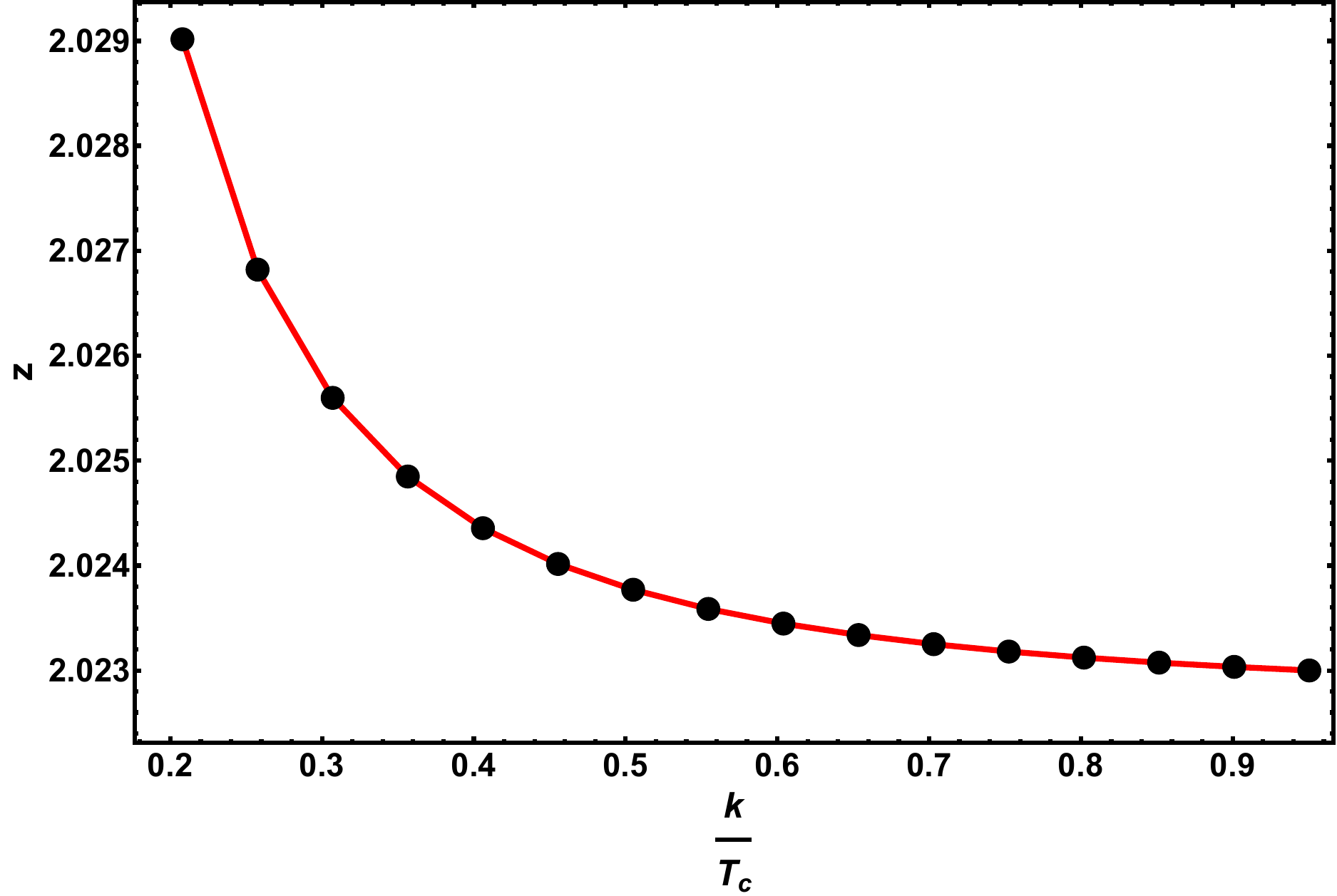}
   \caption{The calculated dynamical critical exponent $z$ by using different spatial momentum. The black circles refer to original data. Although the exact values of $z$ have small deviations from the analytical value $2$, this amount of deviation does not beyond our tolerance.}
   \label{fig:fitz}
\end{figure}

We can directly fit the long-time relaxation behavior according to eq. \ref{eq:3.6}. The discussion above neglect possible effects coming from the new dimensional parameter $k$, that is, the spatial momentum of the corresponding Goldstone modes. Remarkably, the direct calculation shows that the relaxation behavior has very little momentum dependence and the numerical value of dynamical critical exponent $z$ is very close to the desired value $2$, which is referred to model A. In Fig. \ref{fig:fitz}, the exponent $z$ has been fitted from data calculated by using different momentum, ranging from $k/T_c\sim0.2$ to $k/T_c\sim1$. The cases with larger momentum are similar to the no Goldstone case. Considering the possible numerical errors caused by fit, we conclude that the dynamical critical exponent $z$ of our system is $2$ even with Goldstone modes involved, and this means our system should be classified as model A according to \cite{hohenberg1977theory}.

Furthermore, one can generalize scaling hypothesis \ref{eq:3.4} to include possible contribution from initial condition. This generalization introduces a new exponent $x$ and the corresponding scaling hypothesis reads
\begin{align}
    \label{eq:3.7}
    \sigma(\sigma_i,t)=b^{-\beta/\nu}\sigma(\sigma_i b^{x\beta/\nu},tb^{-z}).
\end{align}
Similar to the trick we used before, we can choose the value of scaling factor $b$ such that $\sigma_i b^{x\beta/\nu}\sim\rm{constant}$. Hence, the scaling form of order parameter can be obtained
\begin{align}
    \label{eq:3.8}
    \sigma(\sigma_i,t)=\sigma_i^{1/x}f_{\sigma_i} (t\sigma^{z\nu/x\beta}_i)
\end{align}

\begin{figure}[ht]
    \centering
    \includegraphics[scale=0.45]{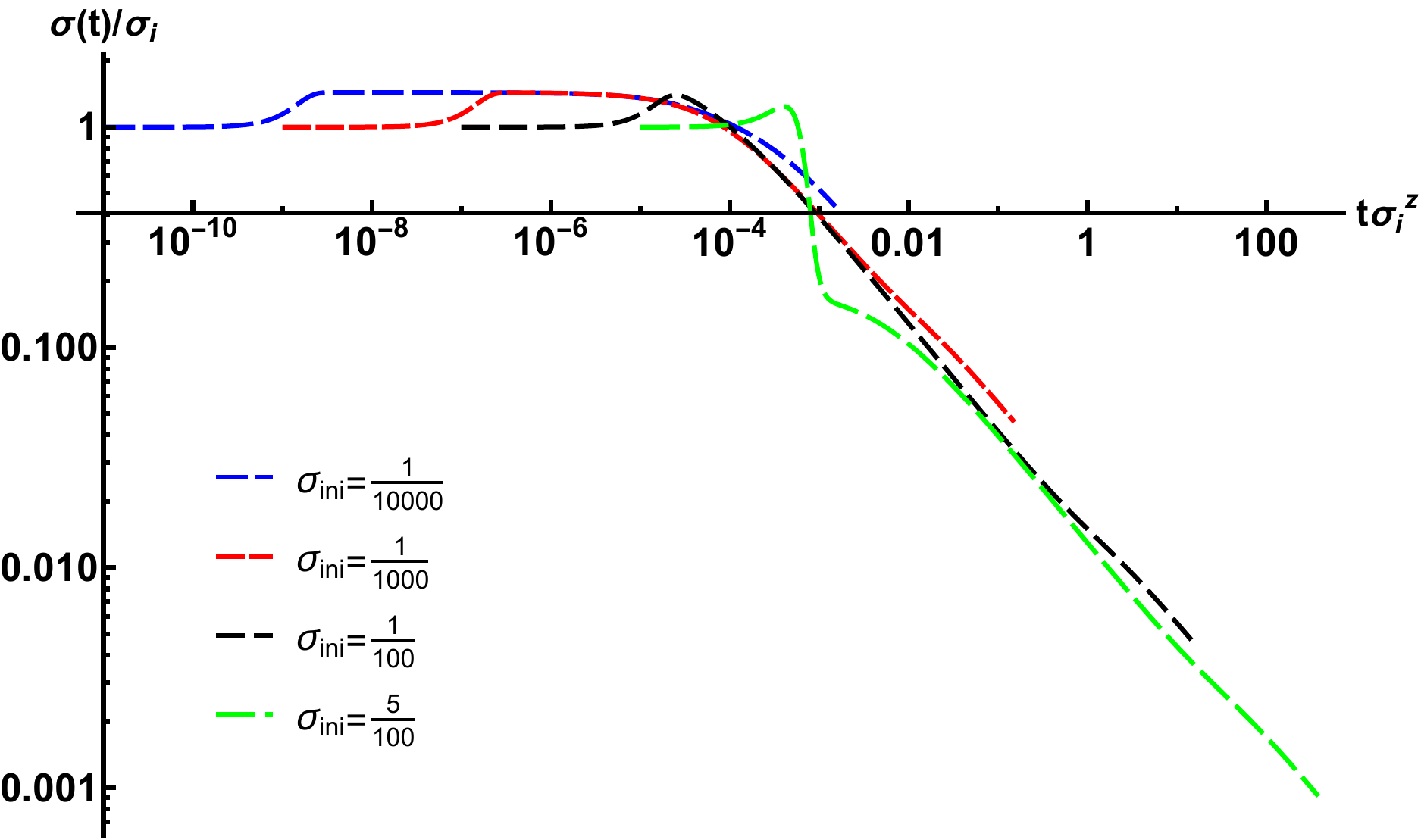}
   \caption{Scaling the evolution curves with different initial conditions based on eq. \ref{eq:3.8}. The spatial momentum of Goldstone modes is fixed to be $k/T_c=1/10$.}
   \label{fig:criScaling}
\end{figure}

For simplicity, we assume the unknown exponent $x$ to be $1$, and check this assumption by direct calculation. Our result is shown in Fig. \ref{fig:criScaling}. In this case, we fix the spatial momentum to be $k/T_c=1/10$ to avoid possible extra influence. It is verified that the simple universal scaling relation \ref{eq:3.8} is not affected by introducing extra Goldstone modes in the system. Moreover, the scaling behavior justifies the assumed value of $x$. Remarkably, this $x$ value is exactly the same as the one obtained from no Goldstone system.  
Besides, a clear scaling behavior is also observed in the intermediate prethermalization stage. 
Compared with the scaling behavior appearing in the no Goldstone system, the Goldstone modes do not modify critical scaling behavior significantly.

\subsubsection{New scaling relation at non-critical temperature}
Analogous to the discussion in \cite{Berges:2015kfa,PineiroOrioli:2015cpb}, we here assume that the order parameter $\sigma$ satisfies the following simple scaling relation
\begin{align}
    \label{eq:3.9}
    \sigma(t,|\mathbf{k}|)=s^{\lambda} \sigma(s^{-z}t,s|\mathbf{k}|),
\end{align}
with a new unknown scaling exponent $\lambda$. In this relation, $z$ still refers to typical dynamical critical exponent and $s$ is the usual scaling factor. We can choose scaling factor $s$ such that $s^{-z} t\sim\rm{const}$. A simple scaling relation can be obtained by using this trick. Its explicit form reads
\begin{align}
    \label{eq:3.10}
    \sigma(t,|\mathbf{k}|)=t^{\lambda/z}f_{\sigma}(t^{1/z}|\mathbf{k}|).
\end{align}

It should be emphasized that the scaling function $f_{\sigma}$ here is totally different from the scaling function $f_{\sigma_i}$, which is discussed in the previous section. This new scaling relation \ref{eq:3.10} has nothing to do with thermal critical point. Instead, it describes the dynamical behavior and manifest fundamental properties of non-equilibrium physics. 

\begin{figure}[ht]
    \centering
    \includegraphics[scale=0.45]{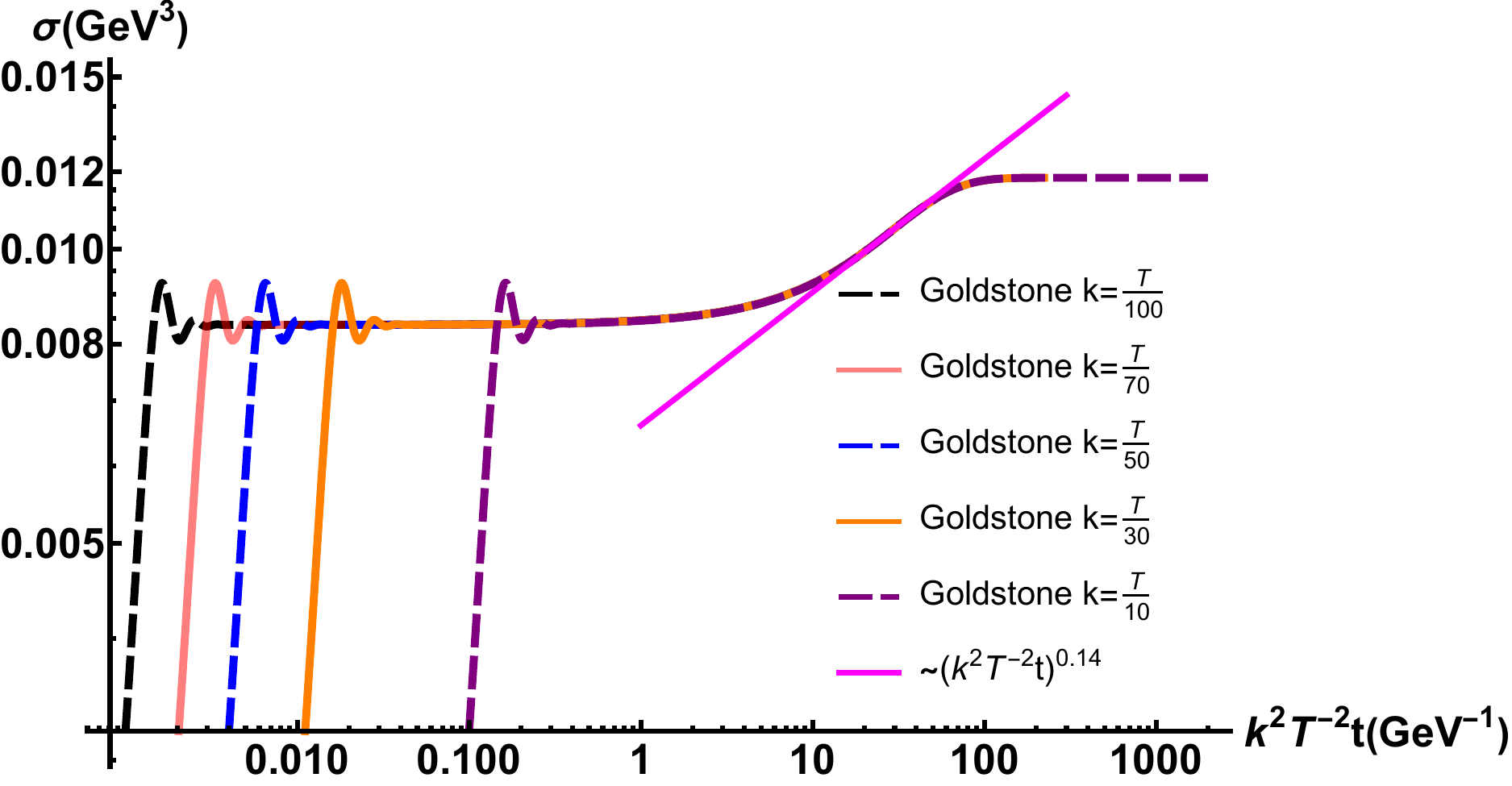}
   \caption{Scaling the evolution curves with different Goldstone momentum at $T=140\rm{MeV}$. The solid magenta line fit the power-law behavior in the relaxation stage. In this scaling form, the relaxation power gives a unique value $0.14$, irrespective of specific momentum. The scaling relation $\sigma(t,k)\propto f(|\mathbf{k}|^2t)$ is an appropriate ansatz and consistent with the assumed scaling relation \ref{eq:3.10}.}
   \label{fig:NoncriScaling}
\end{figure}

Our numerical result is consistent with this scaling assumption. It can be shown directly in Fig. \ref{fig:NoncriScaling} that the different evolution curves, which involve different Goldstone modes, can be well described by a single scaling function $f(|\mathbf{k}|^2t)$. Compared with eq. \ref{eq:3.10}, values of two exponents $z$ and $\lambda$ can be obtained as 
\begin{align}
    \label{eq:3.11}
    z=2,\ \  \lambda=0.
\end{align}
This value of dynamical exponent $z$ is consistent with previous result obtained by studying critical behaviors at exact critical point.   

This new universal scaling behavior can not be some coincidence. Typically, all well-known scaling behaviors arising near critical point can be interpreted by the concept of fixed point. This fact suggests that our new scaling behavior could possibly be related to some fixed point. Note that usual fixed points only reflect thermal equilibrium properties of a given system. On the contrary, the scaling relation introduced in this part reveals dynamical properties of some non-equilibrium system. Thus, we conjecture that non-thermal fixed point should exist in the non-equilibrium evolution and our observed scaling behavior should directly related to it. 

Although in assumption \ref{eq:3.9} we introduce some new exponent $\lambda$, it is irrelevant to our previous discussion according to numerical result. It should be emphasized that the exact physical meaning of $\lambda$ is still missing and need to be further studied.  

\section{Conclusion and discussion}
\label{sec:conclusion}
By using holographic model, we consistently introduce Goldstone modes into our system. The interplay between Goldstone modes and order parameter is naturally encoded in the improved soft-wall AdS/QCD model. By solving this non-linearly coupled system, we can understand some intriguing non-equilibrium properties of nature.

Typically, there exists a quasi-steady state before late-time thermalization when quenching the system from thermal state at critical temperature. We consider this phenomenon as the well-known prethermalization. In our holographic model, this particular dynamical stage can be realized starting from many different initial states. At this stage, the time evolution behavior can be described by some exponential function $\sim e^{-\Gamma t}$ with rather small $\Gamma$. The Goldstone modes with specific spatial momentum have influences on such decay rate $\Gamma$. Consequently, the detailed evolution is modified by such modes.

Most strikingly, we also observe emergence of prethermalization stage even quenching from some non-critical temperature states. The properties of this phenomenon are closely related to momentum of Goldstone modes. With finite non-zero spatial momentum, the period of time of prethermalization is also finite. Subsequently, system evolve into relaxation stage and finally, stay at some thermal equilibrium state. When the Goldstone momentum approaches to zero limit, the duration of prethermalization tends to infinity. This is rather similar to typical critical slowing down. This suggests that some additional fixed point should exist in non-equilibrium evolution. This kind of fixed point can be interpreted as non-thermal fixed point. A new scaling relation which could possibly relate to this conjectured fixed point is also found. Although many evidences of the existence of such fixed point, there remains some flaws in our argument. In particular, from a theoretical viewpoint, we do not find a good way to define number density in our framework. So direct comparison with other methods, including kinetic theory and ordinary Schwinger-Keldysh field theory, is still lacking. Based on this argument, the statement of non-thermal fixed point in our work should be treated as some conjecture.

On the other hand, Goldstone modes do not have an apparent impact on the relaxation stage and final thermalization. In particular, the dynamical critical exponent $z$ which is closely related to critical slowing down phenomenon is not modified. Similar scaling form of evolution curves is also showed. 

The research of non-equilibrium physics is only in its initial phase. Many fundamental questions, either conceptual or technical, are urged to be studied and solved. In recent years, interesting non-equilibrium phenomenon has attracted a lot of attention. We believe that the forthcoming developments of this area will help people better understand our nature.

\begin{acknowledgments}
This work is supported by the National Natural Science Foundation of China (NSFC) Grant Nos: 12305136, 12275108, 12235016, 12221005, 12247107, 12175007,  the start-up funding of Hangzhou Normal University under Grant No. 4245C50223204075, and the Strategic Priority Research Program of Chinese Academy of Sciences under Grant No XDB34030000, and the Fundamental Research Funds for the Central Universities.
\end{acknowledgments}


\bibliographystyle{JHEP}
\bibliography{ref}
\end{document}